\documentclass[fleqn,usenatbib,useAMS]{mnras}

\usepackage{graphicx}	% Including figure files
\usepackage{amsmath}	% Advanced maths commands
\usepackage{amssymb}	% Extra maths symbols
\usepackage{multicol}        % Multi-column entries in tables
\usepackage{bm}		% Bold maths symbols, including upright Greek
\usepackage{pdflscape}	% Landscape pages
\usepackage[usenames]{color}
\definecolor{R}{RGB}{255,0,0}
\definecolor{B}{RGB}{0,0,255}
\definecolor{G}{RGB}{0,150,0}

\usepackage{amsfonts}
\usepackage[T1]{fontenc}
\usepackage{ae,aecompl}

\usepackage{newtxtext,newtxmath}
\usepackage{amsmath}
\usepackage[nopatch]{microtype}
\usepackage{booktabs}
\usepackage{bm}
\usepackage{multirow}
\usepackage{hyperref} 
\usepackage{enumitem}

\title[Morphology and dynamical state of supercluster \textit{cores}]{Nucleation regions in the Large-Scale Structure II: Morphology and dynamical state of supercluster \textit{cores}}

\author[Z\'u\~niga, Caretta \& Andernach]{Johan M. Z\'u\~niga$^1$\thanks{Contact e-mail: 
\href{mailto:jm.zuniga@ugto.mx}{jm.zuniga@ugto.mx}}
, C\'esar A. Caretta$^1$, and Heinz Andernach$^{1,2}$
\\
$^1$Departamento de Astronom\'ia, DCNE-CGT, Universidad de Guanajuato, Guanajuato, GTO, Mexico\\
$^2$Th\"uringer Landessternwarte, Sternwarte 5, D-07778 Tautenburg, Germany}

\date{Last updated 2025 August 1; in original form 2025 April 14}

\pubyear{2025}

\begin{document}
\label{firstpage}
\pagerange{\pageref{firstpage}--\pageref{lastpage}}
\maketitle

\begin{abstract}
This work explores the morphology and dynamical properties of \textit{cores} within rich superclusters, highlighting their role as transitional structures in the large-scale structure of the Universe. Using projected and radial velocity distributions of member galaxies, we identify \textit{cores} as dense structures that, despite being gravitationally bound, are not yet dynamically relaxed. However, they exhibit a tendency toward virialisation, evolving in a self-similar manner to massive galaxy clusters but on a larger scale. Morphological analysis reveals that \textit{cores} are predominantly filamentary, reflecting quasi-linear formation processes consistent with the Zeldovich approximation. Our estimates of the entropy confirm their intermediate dynamical state, with relaxation levels varying across the sample. Mass estimates indicate efficient accretion processes, concentrating matter into gravitationally bound systems. We conclude that \textit{cores} are important environments where galaxy evolution and hierarchical assembly occur, bridging the gap between supercluster-scale structures and virialised clusters.
\end{abstract}


%%%%%%%%%%%%%%%%%%%%%%%%%%%%%%%%%%%%%%%%%%%%%%%%%%

%%%%%%%%%%%%%%%%% BODY OF PAPER %%%%%%%%%%%%%%%%%%

\section{Introduction}
The evolution of structures at various scales in the Universe is understood today through the hierarchical formation model \citep[e.g.,][]{pe1980,pad1993} in a $\Lambda$CDM scenario. The level of substructuring that a galaxy system (cluster or group) or larger structure could exhibit is a consequence of this hierarchical formation process. In fact, the presence and {dynamical significance of substructures (i.e., their prominence and measurable impact on the overall properties) in a particular structure can be considered as an observational estimator of its evolutionary state} \citep[e.g.,][]{am2009}. More dynamically evolved systems, such as virialised rich clusters, exhibit a more regular --- spatial and velocity --- distribution of galaxies with negligible or no substructures \citep[e.g.,][and references therein]{ca2023,zu2024b}. 

On the other hand, superclusters are generally not considered dynamically relaxed and globally gravitationally bound structures \citep[e.g.,][]{oo1983,pea2014,san2023}, presenting clear signs of substructuring at various scales ranging from isolated systems to large filaments and dense ``nodes'' in their intersections \citep[e.g.,][]{ei2007b,ir2020}. Superclusters are considered the youngest structures being formed under gravitational influence and still retain the memory of their formation history \citep[e.g.,][]{ch2013,ei2019}, making them important laboratories for testing cosmological models of structure formation and evolution across different scales \citep[e.g.,][]{ei2021}.

The study of the internal structure of rich superclusters has revealed the presence of `central regions' within most of them \citep[][]{ei2007c,ei2008}. These regions, known as \textit{cores}, are characterized by high density contrasts in both galaxy number and mass. In our previous work \citep[][hereafter Paper I]{zu2024a}, a catalogue of such \textit{cores} was compiled for a sample of rich superclusters. Specifically, they were identified as gravitationally bound structures, comprised of two or more clusters and groups, representing significant cosmological overdensities high enough to suggest they will virialise in the future. A supercluster can host more than one \textit{core} depending on its mass; generally, more massive superclusters tend to host a greater number of \textit{cores}. It is more likely to find multiple \textit{cores} within superclusters with masses $\geq 10^{15}\,\ h_{70}^{-1}\mathcal{M}_\odot$, indicating a more complex and densely populated internal structure in these structures.

In Paper I, \textit{cores} are regarded as nucleation regions because they are zones within superclusters where matter accumulates and condenses, forming denser and more compact structures. In these zones, rich clusters act as `seeds' around which additional matter, whether from galaxies, galaxy systems, or filaments connecting different parts of the supercluster, gathers and aggregates. Essentially, \textit{cores} are focal points of dynamic activity and structure formation within superclusters, where processes of accretion and merging of matter are predominant \citep[e.g.,][]{mar2004}. This makes them important regions for the growth and development of large-scale structures. Moreover, the high density and dynamical interactions within \textit{cores} provide unique environments for studying the interplay between dark and baryonic matter, and the influence of cosmic web filaments, thereby offering valuable insights into the mechanisms driving cosmic structure formation and evolution.

In this work, we focus on studying the dynamical state of \textit{cores} through various approaches, including the analysis of the spatial and velocity distributions of their member galaxies, their morphologies, their entropies, and the estimation of their virial parameters. Our goal is to understand the evolutionary state of these structures, apart from finding out whether they are gravitationally bound. Under the hierarchical formation model, we hypothesise that \textit{cores} might be in an intermediate relaxation state between virialised clusters and the superclusters in which they reside.

Throughout this paper a flat $\Lambda$CDM cosmology is used with the following parameters:  Hubble constant $H_0=70$ $h_{70}$ km s$^{-1}$ Mpc$^{-1}$ with matter density $\Omega_m=0.3$, curvature density $\Omega_k=0$, and dark energy density $\Omega_{\Lambda}=0.7$.

\section{Data}
\subsection{Our sample of \textit{cores}}
For this study, we used the full \textit{core} sample from the \textit{Density-based Core Catalogue} (DCC, Paper I), a catalogue containing a total of 105 \textit{cores} in 53 nearby rich superclusters, with redshifts between 0.02 and 0.15. The \textit{cores} were selected from candidate structures that were initially identified using improved percolation techniques \citep[based on the DBSCAN and FoF algorithms, e.g.,][]{es1996,ber2006} applied to samples of galaxy systems present in the regions of rich superclusters of the \textit{Main SuperCluster Catalogue} \citep[MSCC,][]{chow2014} based on Abell/ACO clusters \citep[][]{ab1958,ab1989}. The selection of \textit{cores} was based on physically motivated density criteria proposed in the literature \citep[see, for example,][]{du2006,ch2015}, defining them as structures with a high probability of future collapse and virialisation despite cosmic expansion.

{
In particular, in Paper I we defined \textit{cores} as galaxy structures with masses $\mathcal{M}\geq 5\times 10^{14}h_{70}^{-1}\mathcal{M}_\odot$, $\mathcal{R}\geq 7.86$ and $\Delta_\text{cr}\geq 1.36$, where}

{
\begin{equation}\label{chon1}
\mathcal{R}\equiv\frac{\rho_\text{ov}}{\rho_\text{b}},
\end{equation}
}

{is the density ratio between the mean mass density of an overdense region, $\rho_\text{ov}$, and the mean local background density, $\rho_\text{b}$, and }  

{
\begin{equation}\label{chon2}
\Delta_\text{cr}\equiv\frac{\rho_\text{ov}}{\rho_\text{cr}}-1,
\end{equation}
}

{
is the density contrast, with $\rho_\text{cr}=3H^2(z)/8\pi G$ being the critical density of the Universe at redshift $z$. These parameters are used to assess whether a given structure is likely to remain gravitationally bound and eventually virialise in the future.
}

The DCC catalogue contains \textit{cores} of some well-known superclusters of the Local Universe (such as \textit{Corona Borealis}, \textit{Shapley}, \textit{Ursa Major}, \textit{Coma-Leo}, \textit{Hercules}, \textit{B\"ootes}, among others), which have already been identified and studied in previous works \citep[e.g.,][]{ko1998,bar1994,ei2008}. Additionally, new \textit{cores} within other superclusters were identified, generating a systematically constructed and statistically significant sample for studying these structures.

For each MSCC supercluster in our sample, we selected galaxy samples with spectroscopic redshift from the Sloan Digital Sky Survey \citep[SDSS DR13,][]{al2017}, the 2dF Galaxy Redshift Survey \citep[2dFGRS,][]{col2001}, or the 6dF Galaxy Survey \citep[6dFGS,][]{jo2009}, depending on the region of the sky where it was located. Galaxy systems were identified from these samples using iterative system identification algorithms based on the works of \citet{bi2006} and \citet{ir2020}. Our full sample of systems consisted of a total of 3337 groups and clusters, including about 527 Abell/ACO clusters, as well as many others that match systems from other published catalogues of galaxy systems. 

{
Beyond the homogeneity in the identification and analysis algorithms (which also include adjustments to local densities), we also limited our samples of superclusters to the ones completely inside the area and restricted to the redshift limits of each survey, as captured in Figure 1 of paper I. SDSS-DR13 contains the Sloan Legacy Survey (DR7), which is complete and deep enough for our aims, despite receiving photometric and spectroscopic improvements through the subsequent DRs. SDSS-DR13 and 2dFGRS have similar depth, which guarantee the coverage to $z_{\rm lim} \sim 0.15$, while 6dFGS is shallower ($z_{\rm lim} \sim 0.08$) and was used only up to this redshift limit.}

\subsection{\textit{Core} galaxy sample}
The member galaxies of each DCC \textit{core} were defined as all galaxies up to a distance of 3.5$R_\text{vir}^i$ from the centroid of each $i$-th member system in the supercluster galaxy sample in 3D rectangular coordinates, and corrected for the corresponding Finger-of-God effect \citep[FoG, \textit{e.g.},][]{co2012}. Following \citet{ir2020}, we refer to this as the ``supercluster box''. The transformations from equatorial to rectangular coordinates were performed in the form

\begin{equation}\label{xyz}
\begin{split}
& X=D_\text{c}\cos{\delta}\cos{\alpha},\\
& Y=D_\text{c}\cos{\delta}\sin{\alpha},\\
& Z=D_\text{c}\sin{\delta},
\end{split}
\end{equation}
where $\alpha=\text{RA}$, $\delta=\text{Dec}$ and
\begin{equation}\label{D_c}
D_\text{c}(z)=\frac{c}{H_0}\int_{0}^{z}\frac{dz'}{E(z')},
\end{equation}
is the line-of-sight comoving distance \citep[\textit{e.g.},][]{hog2000} of the object defined by its redshift $z$, $c$ is the speed of light, and 
\begin{equation}\label{E}
E(z)\equiv\sqrt{\Omega_m(1+z)^3+\Omega_{\Lambda}}.
\end{equation}

Within a distance of 3.5$R_\text{vir}^i$ from the centre of each member system, we expect to include galaxies up to their turn-around zone (a region that encompasses galaxies on the zero-velocity surface of each system, as well as those that are decelerating while approaching this surface, and those that have begun moving toward collapse towards the system's centre), along with galaxies in bridges between them and galaxies in the dispersed component of the \textit{cores}.

\section{Galaxy distributions in \textit{cores}}
A first analysis of the dynamical state of the \textit{cores} was carried out for their selection process (see Paper I), making it clear that they are mostly gravitationally bound structures. Furthermore, since the \textit{cores} represent cosmologically significant overdensities, they could already be in the process of, or close to, gravitational collapse, to then reach virial equilibrium in the future. However, although the marginal state of equilibrium of the \textit{cores} is well known --- or at least theoretically assumed --- due to their self-similarity to any galaxy system, little or nothing is known about their current dynamical states. We do not know how far or close they are from such an equilibrium.

The study of the spatial and velocity distributions of the member galaxies of a system helps to give us an idea of its evolutionary state. In fact, observations reveal that, as the galaxy system approaches dynamical equilibrium, such distributions tend to statistically known shapes \citep[see structure evolution in simulations, e.g.,][]{am2009}. For example, galaxies in regular clusters tend to have a normal (or Gaussian) line-of-sight velocity distribution, as well as a projected (spatial) distribution that is very well fit by a King profile \citep[e.g.,][]{ad1998,sam2014,zu2024b}. Since the dynamics of the galaxy structures that have detached from the Hubble flow is only dominated by gravity, they evolve in a self-similar way to smaller galaxy systems tending towards virialisation. Thus, the spatial and velocity distributions of member galaxies of \textit{cores} are expected to change throughout the evolutionary processes they undergo to reach the dynamical relaxation like in very rich clusters.

\subsection{Velocity distribution of galaxies} \label{vel_dist}
The line-of-sight velocity distributions of galaxies in the \textit{cores} were studied from the spectroscopic redshifts $z_i \approx V_{r_i}/c$ of galaxies, where $V_{r_i}$ is the radial velocity of the $i$-th galaxy and $c$ the speed of light. For each DCC \textit{core}, the set of redshift values of its $N_g$ member galaxies was examined to explore whether it follows an underlying normal distribution. A preliminary analysis of skewness and kurtosis provided an initial qualitative assessment of the redshift distributions, revealing a variety of patterns: some closely resembling normal distributions, while others exhibited features such as flattened shapes, double peaks, or heavy tails. However, this initial exploration was intended only as a guide to identify possible deviations and to gain some insight into the underlying dynamics of the galaxies within the \textit{cores}, as the subsequent analysis focused on quantifying these deviations using more robust statistical tests.

To formally assess the extent of these deviations, we applied a suite of statistical tests, including the Anderson-Darling (AD), Jarque-Bera (JB), and Lilliefors (L) tests. These tests evaluate whether the null hypothesis --- that a sample of redshift values follows a normal distribution --- can be rejected in favor of an alternative hypothesis suggesting significant deviation from normality. Each test returned a binary result, where $h_\text{test} = 1$ indicated rejection of the null hypothesis (when $p$-values $<0.05$), and $h_\text{test} = 0$ otherwise. The results of these tests, presented in Columns 2 and 3 of Table \ref{tab:core_prop}, provide a more rigorous and quantitative evaluation of the observed patterns, allowing us to distinguish between genuine deviations and random fluctuations from normality.  

By combining the results from the three statistical tests, we found that about 30\% of the studied \textit{cores} exhibit line-of-sight velocity distributions consistent with an underlying normal distribution within a significance level of 0.05, while the remaining 58\% deviate from this behavior. Although the results for another 12\% suggest that in some \textit{cores} the radial velocity distributions are not far from normal, this alone is insufficient to conclude that they are dominantly relaxed structures. While a more advanced evolutionary state compared to their host superclusters is probable, the \textit{cores} still remain highly substructured. Since the radial velocity distributions of galaxies in \textit{cores} are mostly non-Gaussian, we infer that these galaxies have not yet reached a Maxwellian\footnote{This arises because each component of the total velocity is treated a statistically independent random variable, and projecting a Maxwellian distribution (typical of systems in equilibrium) onto a single axis results in a Gaussian distribution \citep[e.g.,][]{bt2008}.
} 3D velocity distribution, which indicates that the \textit{cores} are not dynamically relaxed, as expected. This lack of ``relaxation'' in the velocity distributions is a key indicator that \textit{cores} are still in the process of forming and evolving.

\subsubsection{Bimodality in the distribution of velocity dispersions of core galaxies}
\begin{figure} %<left> <lower> <right> <upper> 
\centering
\includegraphics[trim={2.5cm 0cm 3cm 0cm},clip, width=\columnwidth]{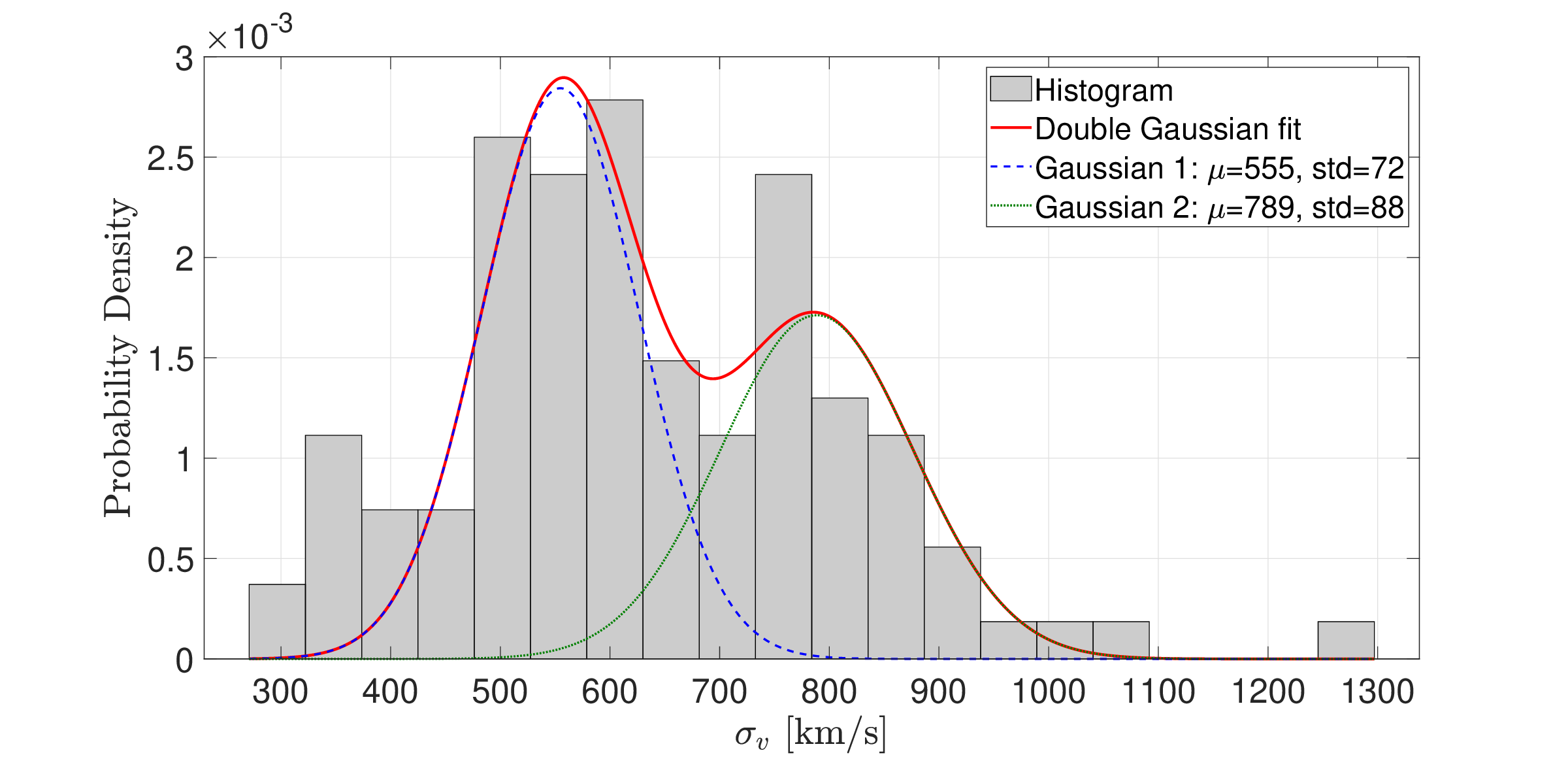} %
 \caption{Distribution of line-of-sight galaxy velocity dispersions in DCC \textit{cores}, along with the double Gaussian fit (red solid line) and its components: Gaussian 1 (blue dashed line) and Gaussian 2 (green dotted line).} 
 \label{f:v_disp}
 \end{figure}

Since the line-of-sight velocity distributions of member galaxies deviate from normality in most \textit{cores}, their bulk velocities and velocity dispersions must be calculated using robust statistical estimators. In particular, here the bulk velocities and velocity dispersions in the line of sight (LOS) were estimated using Tukey's biweight method \citep[e.g.,][]{br1990}. Thus, we take $\bar{V}_{LOS}=C_\text{BI}$ and $\sigma_v=S_\text{BI}$, where $C_\text{BI}$ and $S_\text{BI}$ are, respectively, the robust estimators for the centre and the scale of the velocity distributions of \textit{core} galaxies. The $\bar{V}_{LOS}$ and $\sigma_v$ values obtained for the DCC \textit{cores} are presented respectively in columns 4 and 5 of Table \ref{tab:core_prop}. The distribution of velocity dispersions in the DCC \textit{cores} can be seen in Figure \ref{f:v_disp}. The mean and median values of this distribution are 633 km s$^{-1}$ and 613 km s$^{-1}$, respectively. These galaxy velocity dispersion values are lower than those of typical rich galaxy clusters \citep[$\sim 750-1000$ km s$^{-1}$, e.g.,][]{ba1996,sc2015}.

In virialised systems, the galaxy velocity dispersion $\sigma_v$ directly depends on the virial mass $\mathcal{M}_\mathrm{vir}$: more massive systems generate stronger gravitational potentials, leading to higher orbital velocities and, consequently, larger velocity dispersions for the member galaxies. According to the virial theorem, $U = -2K$, a greater mass implies a deeper gravitational potential well ($U$), which in turns results in a higher internal kinetic energy ($K$), reinforcing this idea \citep[e.g.,][]{zu2024b}.

In non-virialised structures, such as superclusters and their \textit{cores}, the galaxy velocity dispersion $\sigma_v$ serves primarily as a mere statistical measure of the spread in velocities among member galaxies, rather than as a direct indicator of mass. Unlike virialised systems, where $\sigma_v$ is proportional to the total mass, the velocity dispersions in \textit{cores} are not completely tied to their mass. This is because non-virialised structures are out of equilibrium, with their dynamics dominated by ongoing processes such as gravitational collapse or cosmic expansion. As a result, the velocity dispersion can be significantly influenced by these large-scale motions, making it a less reliable tracer of mass in these cases. For instance, the velocity dispersion of an expanding or collapsing structure may primarily reflect the impact of bulk flows associated with individual substructures, rather than representing a single cohesive internal dynamic \citep[e.g.,][]{st1977}.

Note that the $\sigma_v$-distribution in DCC \textit{cores} exhibits a clear bimodal pattern, as seen from the double Gaussian fit (red solid line) shown in Figure \ref{f:v_disp}, with peaks centred around 555 km s$^{-1}$ (Gaussian 1 with blue dashed line) and 789 km s$^{-1}$ (Gaussian 2 with green dotted line). This bimodality remains robust even when varying the binning parameters, suggesting that it is intrinsic to the distribution rather than an artifact of the analysis. To understand the origin of this feature, we consider the role of the most massive cluster (MMC) within each \textit{core}. By definition, each \textit{core} contains one system identified as the MMC. However, the degree to which the MMC dominates the \textit{core}'s dynamics varies significantly.

A separate analysis of the \textit{cores} associated with each Gaussian component confirms this distinction: \textit{cores} associated with the Gaussian 1 (component for low velocity dispersions) typically contain a modest-mass MMC that, while formally the most massive system by the value of its virial mass, is not overwhelmingly dominant. In these cases, the other gravitationally bound systems within each \textit{core} have comparable masses and contribute similarly to the overall internal dynamics of their host \textit{cores}. As a result, the velocity dispersion of \textit{cores} remains relatively low, reflecting a more distributed influence among multiple systems. Conversely, \textit{cores} falling within the Gaussian 2 (component for high velocity dispersions) tend to host an MMC with a significantly larger mass than the other \textit{core} members. Here, the velocity dispersion of galaxies within the \textit{core} is strongly influenced by the MMC, resulting in values characteristic of cluster-like systems, where the virialised portion dominates. Furthermore, since \textit{cores} are collapsing structures, their velocity dispersions can be largely influenced by the bulk motions of their member systems \citep[e.g.,][]{st1977}.

On the other hand, we also observe that a few \textit{cores} contain two or even three highly massive systems with comparable masses, making it reasonable to consider that they collectively dominate the dynamics of the host \textit{cores}. Interestingly, these cases also tend to fall within the Gaussian 2, likely due to the high velocity dispersions of the massive member clusters. This reinforces the assumption that the bimodal velocity dispersion distribution of the DCC \textit{cores} is primarily driven by the varying degrees of dynamical influence exerted by the most massive systems within each \textit{core}.

 \subsection{Projected distribution of galaxies} \label{sp_dist}
To study the spatial distribution of galaxies, we analyzed their projected positions in the plane of the sky to assess whether they exhibit a statistical tendency to follow the projected King profile:

\begin{equation}\label{K_2D}
\Sigma(r)=\Sigma_0\left[1+\left(\frac{r}{r_c} \right)^2 \right]^{-\gamma},
\end{equation}  

where the parameters $r_c$, $\Sigma_0$, and $\gamma$ are determined by fitting the model \eqref{K_2D} to the projected distribution of member galaxies in each \textit{core}.

The King profile was selected because, despite being a mass density profile, its projected form provides a good fit to the radial distribution of galaxies in regular clusters \citep[e.g.,][]{ro1972,ad1998}. These clusters typically feature a dense core of galaxies surrounded by a sparse halo, where the galaxy number density decreases with distance. In contrast, other profiles, such as the Einasto and NFW profiles \citep[e.g.,][]{ei1965,nfw1996}, are better suited to describe the distribution of dark matter within clusters.

Since the MMCs represent the most significant gravitational potential wells in their respective host \textit{cores}, we assume these clusters to be the main physical centres of gravitational attraction within the \textit{cores}. {Although in some \textit{cores} (particularly those in Gaussian 1) a clearly dominant MMC is not present, we adopt the coordinates of the most massive cluster as a reference `centre'\footnote{It is important to mention that studies of the spatial distribution of galaxies in a structure (or system), as well as estimates of certain dynamical parameters, are sensitive to the choice of a `centre'. {While other definitions such as the centre of mass may provide valuable complementary insights, we defer such analysis to future work.}} for consistency across the sample. MMCs, even when not overwhelmingly dominant in mass, are generally rich systems located in high-density regions, suggesting a strong gravitational influence and a likely site of future collapse. Moreover, they often coincide with local peaks in the spatial distribution of galaxies.}

Thus, taking the sky coordinates of each \textit{core}'s MMC as its centre, we calculated the projected \textit{core}-centric distances (in units of $h_{70}^{-1}$ Mpc) of its member galaxies as follows:
\begin{equation}\label{d_core}
R_i\simeq \frac{\pi}{180}\frac{D_\text{c}(\bar{z})}{(1+\bar{z})} \left[(\alpha_{\text{MMC}}-\alpha_i)^2\cos^2{\bar{\delta}}+(\delta_{\text{MMC}}-\delta_i)^2\right]^{1/2},
\end{equation}
where $\alpha_{\text{MMC}}$ and $\delta_{\text{MMC}}$ are the RA and Dec coordinates of the MMC (see columns 6 and 7 of Table 4 in Paper I), $\alpha_i$ and $\delta_i$ are the coordinates of each member galaxy, and $\bar{z}$ and $\bar{\delta}$ represent the mean redshift and declination of the \textit{core} galaxies, respectively. As before, $D_\text{c}(\bar{z})$ is the comoving distance at redshift $\bar{z}$, corresponding to the mean radial distance of each \textit{core}.

We used a binning method to analyze the projected density distribution of galaxies as a function of the \textit{core}-centric distance in the RA-Dec plane. This process involved counting only member galaxies in concentric circular annuli in 2D, centred on the MMC of the given \textit{core}. Each ring was taken to have an area $A_r = 2\pi R \Delta R$, where $R$ is the mean radius of the ring and $\Delta R$ is its width. The width $\Delta R$ was kept constant for all radii, and all galaxies with \textit{core}-centric distances $R_i$ between $R - \Delta R/2$ and $R + \Delta R/2$ were counted within the corresponding ring. The surface density of galaxies at a given radius $R$ was then calculated as $\Sigma(R) = N_r/A_r$, where $N_r$ is the number of galaxies in that ring.

\begin{figure} %<left> <lower> <right> <upper> 
\centering 
\includegraphics[trim={10cm 0cm 10cm 1cm},clip, scale=0.4]{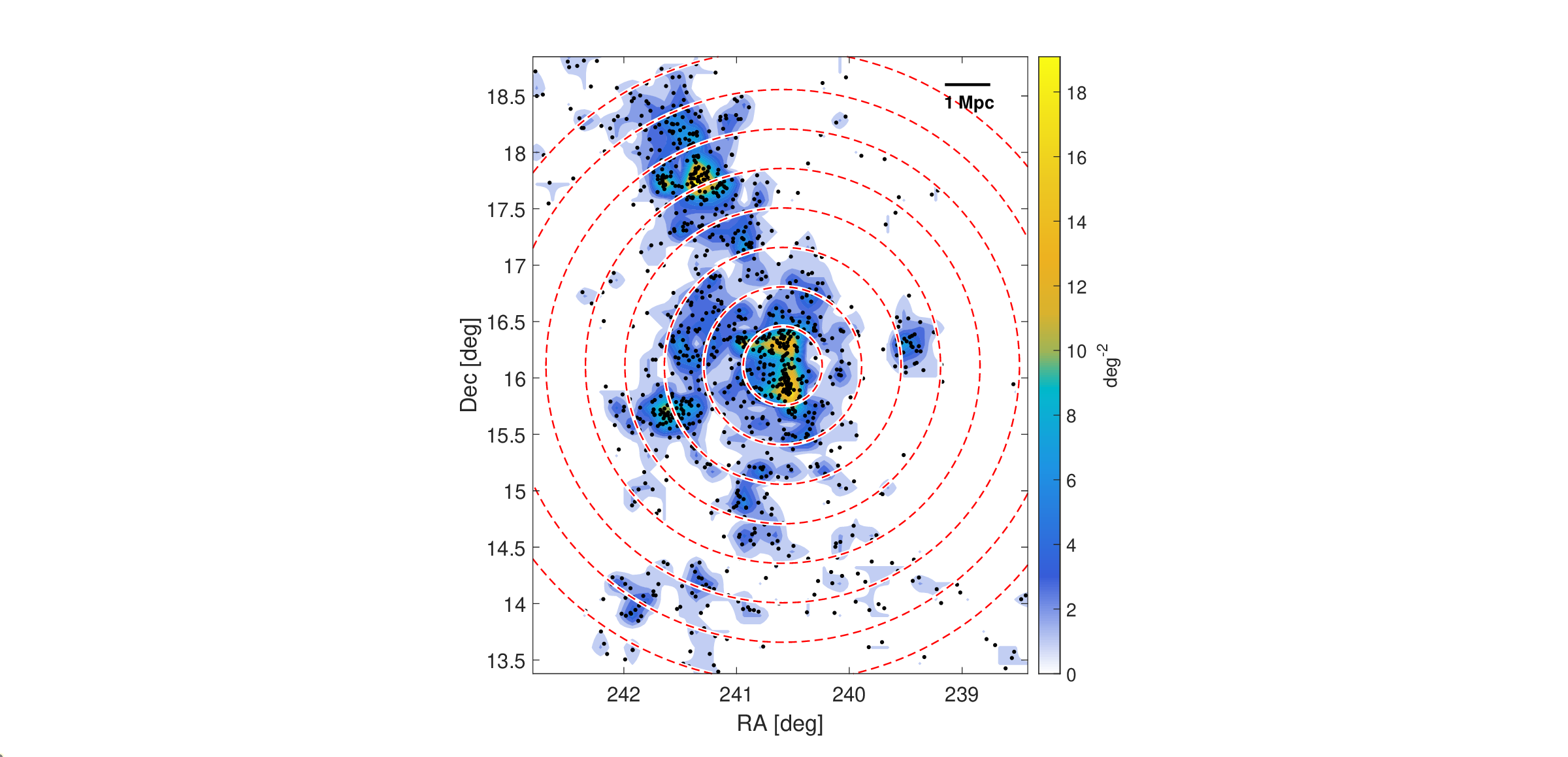} \\
\includegraphics[trim={2.5cm 0cm 2.5cm 0.5cm},clip,width=\columnwidth]{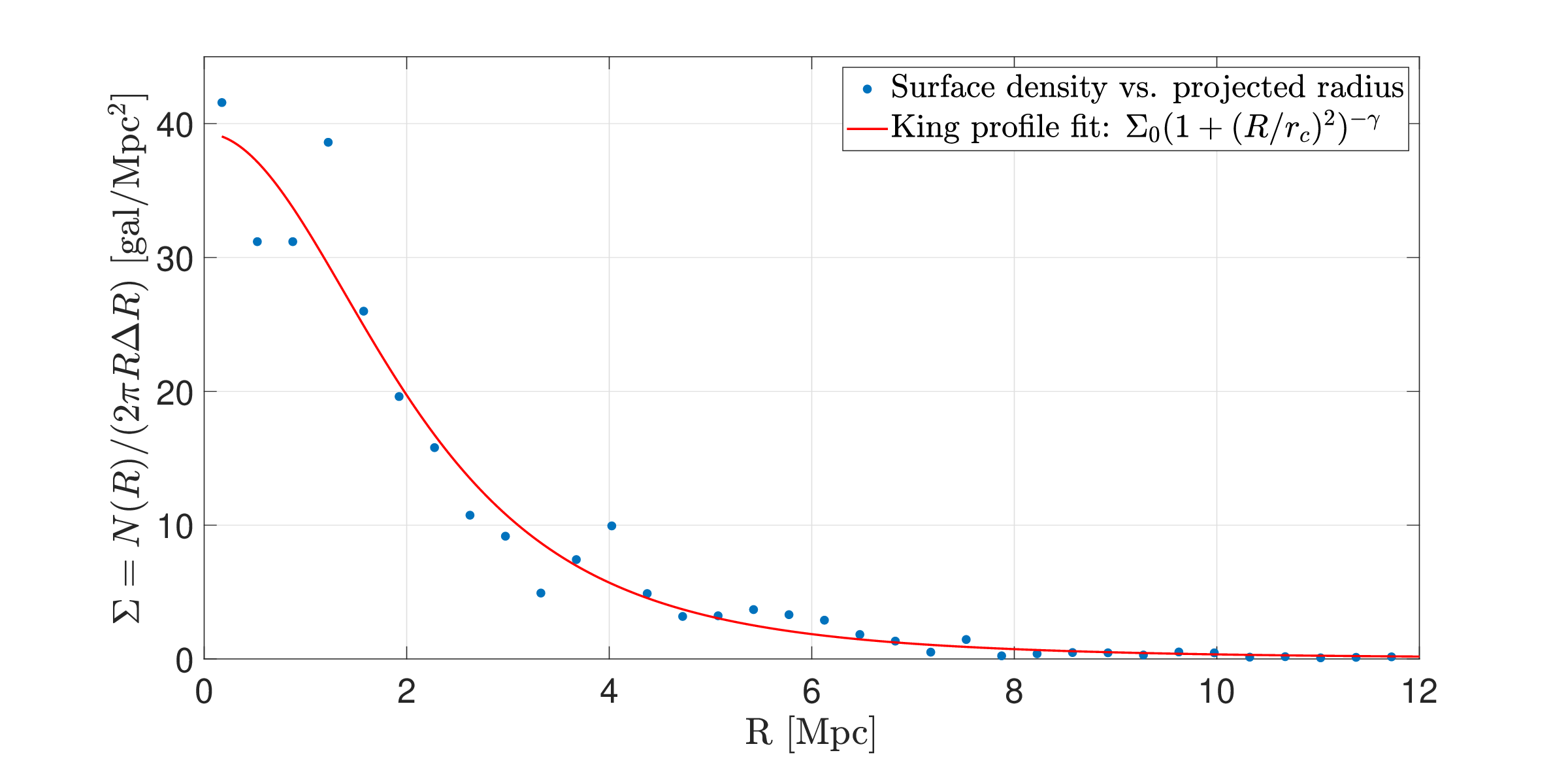} \\
 \caption[]{\textit{Top}: 2D-density map of the projected distribution of galaxies in the DCC 099 \textit{core} of the \textit{Hercules Supercluster} (MSCC 474). Each black dot represents a member galaxy of the \textit{core}. The space between the red dashed circles give a schematic representation of the rings (bins in 2D) on which galaxies were counted ($\Delta R=0.35$ $h_{70}^{-1}$Mpc). The width of the annuli plotted was chosen as 1 $h_{70}^{-1}$Mpc to avoid saturation of circles in the graph. \textit{Bottom:} $\Sigma$ \textit{vs.} $R$ plot for the DCC 099 \textit{core}. The blue dots represent the surface number density of galaxies --- in the ring --- at distance $R$ from the centre (the MMC) of the \textit{core}. The solid red line is the projected King profile fitted to the data using the NLS method. The fit parameters for this case were $\Sigma_0=44.68$ gal/$h_{70}^{-2}$Mpc$^{2}$, $r_c=2.56$ $h_{70}^{-1}$Mpc and $\gamma=1.78$, with a goodness of fit $R_\text{det}^2=0.91$.} 
 \label{f:flucc103}
 \end{figure}

For all \textit{cores}, we adopted a fixed ring width of $\Delta R = 0.35$ $h_{70}^{-1}$ Mpc (a value that maximized the goodness-of-fit in most \textit{cores}), while allowing the radius $R$ to vary without overlap between bins. Using the Nonlinear Least Squares (NLS) method, we fitted the projected King profile \eqref{K_2D} to the resulting set of $(R, \Sigma)$ pairs, determining the best-fit parameters $(\Sigma_0, r_c, \gamma)$. Figure \ref{f:flucc103} provides an example of this procedure applied to the DCC 099 \textit{core} in the \textit{Hercules Supercluster} (MSCC 474). A similar analysis was performed for all other \textit{cores}.

Columns 6 to 9 of Table \ref{tab:core_prop} present the set of parameters $(\Sigma_0,r_c,\gamma)$ obtained by fitting the King profile to the projected galaxy sample of each DCC \textit{core}, along with the respective goodness of fit measured through the $\mathcal{R}^2_\text{det}$ statistic\footnote{Also known as the coefficient of determination. This statistic assesses the quality of a model in replicating results and indicates the proportion of variation in the results that can be explained by the fitted model.}. The profile fitting using the NLS method was not successful for approximately 33\% of the DCC \textit{cores}, resulting in unphysical parameters that were excluded from the analysis. The mean and median values of the corresponding distributions of parameters $(\Sigma_0,r_c,\gamma)$ in the sample of DCC \textit{cores}, excluding outliers, are presented in Table \ref{tab:king_p}. Notably, the characteristic radius $r_c$ of the \textit{cores}, with a median value of $1.93$ $h_{70}^{-1}$Mpc, is comparable to the virial radii of their MMCs \citep[in contrast to the median values for galaxy clusters of about 0.25 $h_{70}^{-1}$Mpc, e.g.,][]{ba1996,ca2023}. This may suggest that, although DCC \textit{cores} are not yet regular galaxy structures, they have the potential to evolve into `core-halo supersystems', with their central nuclei likely located close to their MMCs. It can be expected that these systems will serve as the formation regions for the future nuclei of marginally virialised (relaxed and regular) \textit{cores}.

 \begin{table}
\caption[]{Mean (with standard deviation) and median values for the $\Sigma_0$, $r_c$ and $\gamma$ parameters (excluding outliers\footnote{{Throughout this work, outliers were identified using the standard interquartile range (IQR) rule, i.e., data points outside $[Q_1-1.5\times \rm{IQR}, Q_3+1.5\times \rm{IQR}]$ were excluded from the analysis.}}) of the King profile fit to the projected distributions of galaxies in the sample of DCC \textit{cores}. Median values are given as asymmetric ranges with $\Delta\text{Q}_1=\mathrm{Median}-\text{Q}_1$ and $\Delta\text{Q}_3=\text{Q}_3-\mathrm{Median}$ as lower and upper indices, where $\text{Q}_1$ and $\text{Q}_3$ are the 25th and 75th percentiles, respectively.}

\label{tab:king_p}
\centering %
\resizebox{6.5cm}{!}{
\begin{tabular}{crr}
\hline\hline
Parameter  & Mean $\pm$ std & Median$_{-\Delta\text{Q}_1}^{+\Delta\text{Q}_3}$ \\ \hline
$\Sigma_0$ [gal/$h_{70}^{-2}$Mpc$^2$] & $10.24 \pm 7.69$ & $7.38_{-2.89}^{+7.50}$ \\
$r_c$ [$h_{70}^{-1}$Mpc]     & $2.17 \pm 1.04$   &  $1.93_{-0.57}^{+0.81}$\\
$\gamma$   & $1.93 \pm 0.99$   &  $1.64_{-0.48}^{+0.78}$ \\
\hline
\end{tabular}}
\end{table}

Approximately 55\% of the DCC \textit{cores} could be fitted by the King profile achieving a goodness of fit $\mathcal{R}^2_\text{det} > 0.9$, while only $\sim$8\% of them were fitted with a goodness of fit $\mathcal{R}^2_\text{det} < 0.8$. Thus, most of the \textit{cores} in the sample exhibit a projected distribution of galaxies that can be adequately described by a King profile, suggesting an evolutionary tendency toward dynamical states statistically consistent with this density model. However, the observed density distributions still display ``humps'' as a function of radius, indicating the presence of substructures within the \textit{cores}. These humps can affect the values of the fit parameters, which, in principle, assuming that the distribution of galaxies follows the distribution of total matter (dark and baryonic) and \textit{vice versa} in these structures, should be related to the equilibrium state of the \textit{cores}. For instance, and as anticipated, the $\gamma$ values obtained for the distributions of \textit{core} galaxies remain far from $\gamma = 1$, which is the expected terminal value for a projected galaxy distribution in dynamical equilibrium \citep[e.g.,][]{sa1986,ad1998}.

\section{Morphology analysis of \textit{cores}}\label{morph}
Studying the morphology of galaxy structures also provides valuable insights into their dynamical states or evolutionary phases. The shape of a structure is primarily influenced by the distribution of matter within it and the gravitational interactions among its member galaxies and galaxy systems. Consequently, the morphology of these structures can reveal information about the spatial distribution of galaxies and dark matter within them, as well as their interactions with one another. Additionally, the shape may be related to their formation and evolutionary history, and it can be influenced by external factors, such as interactions with other large-scale structures in the Universe. \\

Since \textit{cores} are not virialised structures, they exhibit a wide variety of shapes. In this section, a morphometric analysis of the DCC \textit{cores} will be conducted using statistical and geometrical methods.

\subsection{Minkowski Functionals and Shapefinders}\label{mk}
A comprehensive morphological study of bodies in \(n\) dimensions requires both topological and geometrical descriptors to characterise their connectivity, content, and shape \citep[e.g.,][]{mec1994}. Minkowski functionals (MFs) constitute a family of \(n+1\) morphological descriptors for extended bodies, grounded in well-established principles of integral geometry \citep[e.g.,][]{wi2014}. The morphological properties of an \(n\)-dimensional body are determined by the \((n-1)\)-dimensional hypersurface that encloses it \citep[e.g.,][]{she2003}. Thus, the morphology of a closed two-dimensional surface embedded in three-dimensional Euclidean space is comprehensively characterized by four MFs \citep[e.g.,][]{ei2007a,bag2020}:

\begin{enumerate}
    \item The volume enclosed by the surface: \(V\),
    \item The surface area: \(S\),
    \item The integrated mean curvature %\footnote{This is obtained by integrating the mean curvature over the entire surface. The mean curvature at a point on a surface is the average \((\kappa_1+\kappa_2)/2\) of the principal curvatures at that point, where \(\kappa_i=1/R_i\) and \(R_i\) is a radius of curvature defined in the plane normal to the surface at that point. Mathematically, the principal curvatures \(\kappa_1\) and \(\kappa_2\) are the eigenvalues of the curvature matrix (or curvature tensor) at a given point, representing intrinsic local properties of the surface. They depend solely on the geometry of the surface at that specific point and are independent of the location from which the radius is drawn, the coordinate system, or the orientation of the surface in space.}
    of the surface:
    \begin{equation}\label{imc}
    C = \frac{1}{2} \oint \left( \frac{1}{R_1} + \frac{1}{R_2} \right) dS,
    \end{equation}
    \item The integrated Gaussian curvature (or Euler characteristic) of the surface:
    \begin{equation}\label{gc}
    \chi = \frac{1}{2\pi} \oint \frac{1}{R_1 R_2} dS,
    \end{equation}
\end{enumerate}

where in Equations \eqref{imc} and \eqref{gc}, \(R_1\) and \(R_2\) denote the two local principal radii of curvature at any point on the surface. Furthermore, the Euler characteristic can be expressed in terms of the genus, which quantifies the number of topological handles that the surface exhibits and provides a measure of the connectivity of the structure, distinguishing between isolated underdense regions (voids) and interconnected features \citep{she2003}. Essentially, it describes the number of holes in a close surface or three-dimensional object and can be defined as follows: 
\begin{equation}
\mathcal{G} = 1 - \frac{\chi}{2}.
\end{equation} 

Both \(\chi\) and \(\mathcal{G}\) serve as measures of the surface topology \citep[multiply connected surfaces have \(\mathcal{G}>0\), while those that are simply connected have \(\mathcal{G}=0\), e.g.,][]{sah1998}. Thus, while the genus provides insight into the connectivity of a surface, the other three MFs are sensitive to local surface deformations, effectively characterizing the geometry and shape of the bodies \citep{she2003}.

To characterise the shape and determine the characteristic dimensions of a structure, we employ ``shapefinders'' \citep[][]{sah1998}, which are defined from the MFs as follows \citep[e.g.,][]{she2003,bag2020}:

\begin{enumerate}[itemsep=3pt]
\item Thickness: 
\begin{equation}\label{Ts}
\mathcal{T}=3V/S,
\end{equation}
\item Breadth: 
\begin{equation}\label{Bs}
\mathcal{B}=S/C,
\end{equation}
\item Length:
\begin{equation}\label{Ls}
\mathcal{L}=\frac{C}{4\pi(1+|\mathcal{G}|)},
\end{equation}
\item Planarity: 
\begin{equation}\label{Ps}
\mathcal{P}=\frac{\mathcal{B}-\mathcal{T}}{\mathcal{B}+\mathcal{T}},
\end{equation}
\item Filamentarity: 
\begin{equation}\label{Fs}
\mathcal{F}=\frac{\mathcal{L}-\mathcal{B}}{\mathcal{L}+\mathcal{B}},
\end{equation}
\end{enumerate}
This set of five shapefinders includes three quantities with dimensions of length (\(\mathcal{T}\), \(\mathcal{B}\), and \(\mathcal{L}\)) and two dimensionless ratios (\(\mathcal{P}\) and \(\mathcal{F}\)), providing a robust framework for quantifying the morphology of structures.

The shapefinders \(\mathcal{T}\), \(\mathcal{B}\), and \(\mathcal{L}\) estimate the three principal physical extensions of a three-dimensional structure. These shapefinders are spherically normalized, ensuring that \( V = (4\pi/3) \mathcal{T} \mathcal{B} \mathcal{L} \), where \( V \) is the volume enclosed by the structure. Generally, for a convex surface, the relation \( \mathcal{T} \leq \mathcal{B} \leq \mathcal{L} \) holds; if not, the smallest dimension is designated as \(\mathcal{T}\) and the largest as \(\mathcal{L}\) to maintain order \citep{bag2020}. In cases where \( C < 0 \), it is possible to redefine \( C \to |C| \) to ensure that both \(\mathcal{B}\) and \(\mathcal{L}\) remain positive. Oblate ellipsoids (pancake-like shapes) are characterized by \(\mathcal{T} < \mathcal{B} \approx \mathcal{L}\), while prolate ellipsoids (filamentary structures) are described by \(\mathcal{T} \approx \mathcal{B} < \mathcal{L}\) \citep{ei2007a}.

On the other hand, the shapefinders $\mathcal{P}$ and $\mathcal{F}$ are dimensionless quantities that allow us to determine the shape of an object. For example, in some works \citep[e.g.,][]{sah1998,ei2007a,bag2020} these shapefinders have been characterized so that:
\begin{itemize}[itemsep=0pt]
\item For spheres: $\mathcal{T}=\mathcal{B}=\mathcal{L}$, i.e., $\mathcal{P}=\mathcal{F}=0$,
\item For ideal filaments $\mathcal{P}\approx 0$, $\mathcal{F}\approx 1$,
\item For real filaments: $\mathcal{F}\gg\mathcal{P}$,
\item For ideal pancakes: $\mathcal{P}\approx 1$, $\mathcal{F}\approx 0$,
\item For planar objects (sheets or pancakes): $\mathcal{P}\gg\mathcal{F}$,
\item For ideal ribbons: $\mathcal{P}\approx\mathcal{F}\approx 1$,
\item For ribbon-like objects: $\mathcal{P}/\mathcal{F}\approx 1$. 
\end{itemize}
Note that in this context, the word `ideal' refers to a theoretical or simplified representation of the structure, as commonly used in the literature.

\subsection{Morphometry of \textit{cores}} \label{poly_surf}
The MFs are typically defined for extended bodies with well-established (smooth and differentiable) boundary surfaces. However, they can also be applied to galaxy distributions by constructing an extended object from the point set of galaxy coordinates. This is achieved by defining a limiting surface that encloses the member galaxies of a structure, thus allowing its morphology to be characterized by the four MFs ($V$, $S$, $C$, $\chi$) that describe the enclosing surface \citep[e.g.,][]{ei2007a,ei2008}. Given the global and additive nature of MFs, their applicability can be extended to surfaces with singular edges and corners \citep{mec1994}. This versatility makes them suitable for analyzing irregular or discontinuous structures, providing valuable insights into the geometry and topology of \textit{cores}. In this work, we employ two methods to generate enclosing surfaces around the \textit{core} member galaxies, adjusting the definition of the MFs (and shapefinders) in each case to match the specific characteristics of the generated surfaces.

\subsubsection{Main method: polyhedral surfaces} 
\begin{figure*} %<left> <lower> <right> <upper> 
\centering 
\includegraphics[trim={0.5cm 1cm 0.5cm 2cm},clip,width=0.9\textwidth]{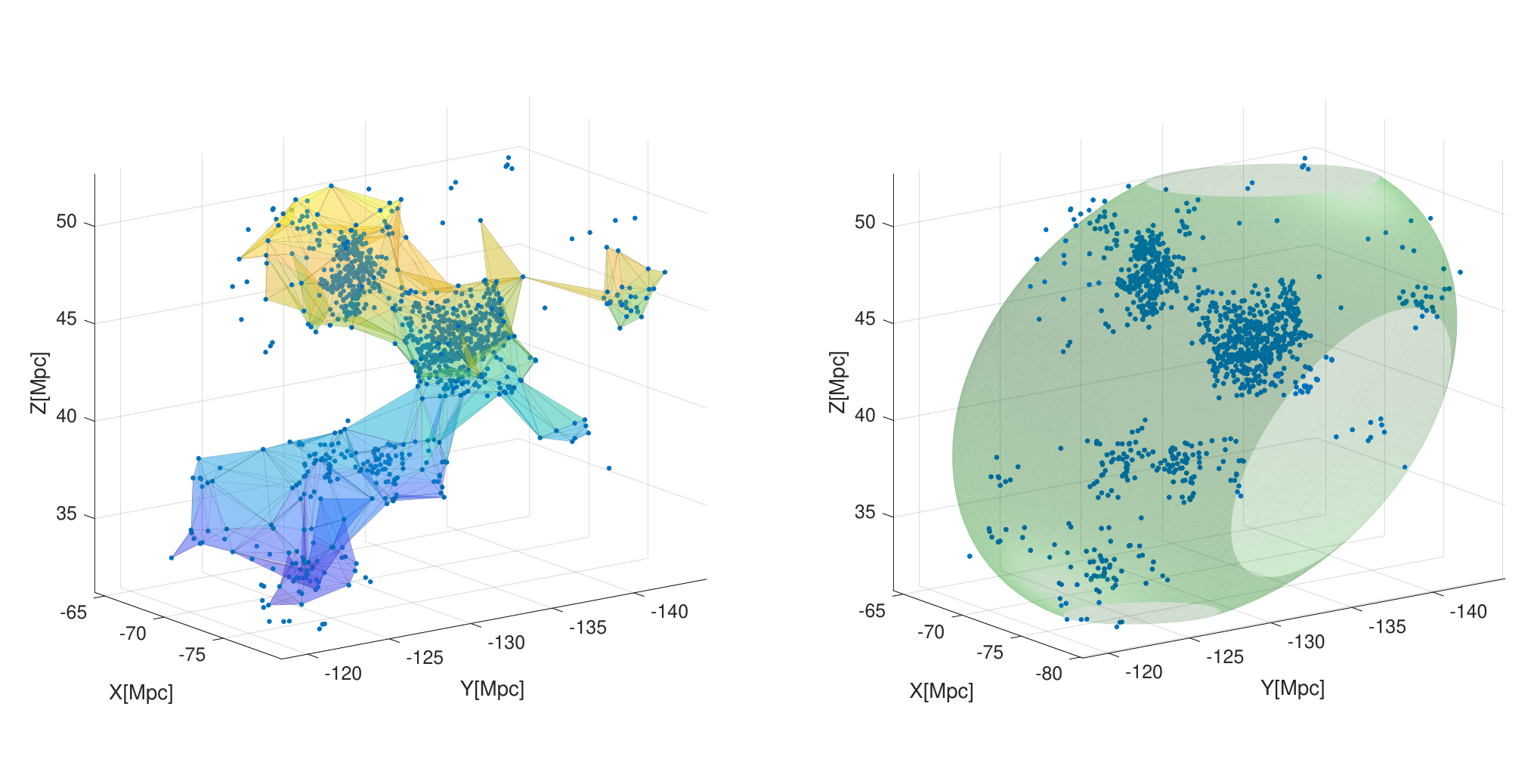} \\
 \caption[]{\textit{Left}: Polyhedral fit of the optimal surface (with shrink factor, $s_\text{f}=1$) enclosing the distribution of sampled member galaxies in DCC 099, the main \textit{core} (A2147-A2151-A2152A-A2153A) of the \textit{Hercules Supercluster} (MSCC 474). \textit{Right}: The alternative ellipsoidal fit for the same galaxy distribution. The lighter-colored regions on the ellipsoid correspond to automatically generated cut planes in the upper XY and right YZ sections, enhancing the 3D visualisation of the surface. {The polyhedral and ellipsoidal surfaces shown are best-fit models of the main overdense body of the\textit{core}, not a strict envelope of all its member galaxies. Due to the complex and unrelaxed nature of the structures, some galaxies, particularly those in the more diffuse or peripheral regions, may fall outside these fitted surfaces.}} 
 \label{f:core_sF}
 \end{figure*}

Polyhedral surfaces that envelop the DCC \textit{cores} were constructed by triangulating boundary points from the three-dimensional distribution of their member galaxies. The triangulation was performed using the 3D alpha-shape algorithm \citep[e.g.,][]{em1994}, implemented through the MATLAB \texttt{boundary} function. {This function identifies and returns the set of boundary points of a distribution for a given compactness level, modulated by the shrink factor $s_\text{f}$, and subsequently allows the generation of a polyhedral envelope, ranging from the \textit{convex hull}, for $s_\text{f}=0$, to a tightly fitted \textit{compact boundary}, for $s_\text{f}=1$ \citep[see][]{mw2022}}. The left panel of Figure \ref{f:core_sF} illustrates an example of the polyhedral surface fitted to member galaxies in the main \textit{core} (DCC 099) of the \textit{Hercules Supercluster} (MSCC 474).

The obtained triangulated polyhedral surface defines a single well-defined boundary around the member galaxies. Morphological properties of these surfaces, including their geometric and topological characteristics, are determined by the MFs adapted for triangulated surfaces \citep[e.g.,][]{she2003,bag2020}:

\begin{itemize}
    \item The total surface area is computed as  
    \begin{equation}
        S = \sum_{k=1}^{N_T} S_k,
    \end{equation}
 
    where $S_k$ is the area of the $k$-th triangle, and $N_T$ is the total number of triangles forming the surface.

    \item The volume enclosed by the surface is the sum of contributions from $N_T$ tetrahedra:  
    \begin{equation}
            V = \sum_{k=1}^{N_T} V_k, \quad V_k = \frac{1}{3} S_k (\bm{n}_k \cdot \bm{P}_k),
    \end{equation}
 
    where $V_k$ is the volume of the $k$-th tetrahedron with base $S_k$ in the $k$-th triangle, with normal vector $\bm{n}_k$ and centroid vector position $\bm{P}_k$, and apex in an arbitrary origin. Note that the sums of $S$ and $V$ are maximum for $s_\text{f}=0$ and minimum for $s_\text{f}=1$.

    \item The integrated mean curvature is calculated as  
    \begin{equation}
            C = \frac{1}{2} \sum_{j,k} l_{jk} \phi_{jk} \epsilon,
    \end{equation}

    where $l_{jk}$ is the length of the common edge between adjacent triangles $j$ and $k$, $\phi_{jk}$ is the angle between their normals, and $\epsilon = 1$ for convex and $\epsilon = -1$ for concave edges \citep{she2003}.

    \item The Euler Characteristic and Genus are defined as  
    \begin{equation}
            \chi = N_T - N_E + N_V, \quad \mathcal{G} = 1 - \chi / 2,
    \end{equation}

    where $N_E$ and $N_V$ are the number of edges and vertices, respectively.
\end{itemize}
This approach ensures a detailed and versatile morphological analysis of \textit{cores}, enabling the characterisation of both their geometry and topology.

\subsubsection{Alternative Method: Ellipsoidal Fitting}\label{ellips}

To validate the polyhedral surface method described above, continuous ellipsoidal surfaces, whose MFs can be calculated analytically, were fitted to the same galaxy samples for each \textit{core}. The algorithm proposed by \citet{py2015} was used to estimate the semi-major axes ($a$, $b$, $c$) of the ellipsoid that best represents the 3D galaxy distribution in rectangular coordinates. {The fitting process was performed iteratively by extracting boundary points using the \texttt{boundary} function for different values of the shrink factor $(0\leq s_\text{f}\leq 1)$. For each $s_\text{f}$, a new set of boundary points was obtained and used to fit an ellipsoidal surface. The optimal ellipsoid was then selected based on the maximum goodness of fit, measured by the Chi-square statistic}.

The right panel of Figure \ref{f:core_sF} illustrates an example of the best ellipsoidal surface fit for the DCC 099 \textit{core}. The parameters of the best fits for each \textit{core} are presented in columns 2 to 5 of Table \ref{tab:core_ell}. The parametric equation for an ellipsoid with semi-major axes $a$, $b$, and $c$ is given by:
\begin{equation}
\bm{r}(\theta, \phi) = a (\sin{\theta}\cos{\phi})\bm{i} + b (\sin{\theta}\sin{\phi})\bm{j} + c (\cos{\theta})\bm{k},
\end{equation}
where $0 \leq \phi \leq 2\pi$ and $0 \leq \theta \leq \pi$.

The four MFs for an ellipsoidal body are expressed as follows \citep[e.g.,][]{lp1969,sah1998}:
\begin{equation}
V = \frac{4}{3}\pi abc,
\end{equation}
\begin{equation}
S = \oint \sqrt{EG - F^2} \, d\theta \, d\phi,
\end{equation}
\begin{equation}
C = \frac{1}{2} \oint \left[\frac{EN + GL - 2FM}{EG - F^2}\right] dS,
\end{equation}
\begin{equation}
\chi = \frac{1}{2\pi} \oint \left[\frac{LN - M^2}{EG - F^2}\right] dS,
\end{equation}
where:
\begin{align*}
    E &= \bm{r}_\theta \cdot \bm{r}_\theta, & 
    F &= \bm{r}_\theta \cdot \bm{r}_\phi, &
    G &= \bm{r}_\phi \cdot \bm{r}_\phi, \\
    L &= \bm{r}_{\theta\theta} \cdot \bm{n}, & 
    M &= \bm{r}_{\theta\phi} \cdot \bm{n}, & 
    N &= \bm{r}_{\phi\phi} \cdot \bm{n},
\end{align*}
$\bm{n}$ is the unit normal vector to the surface at any point, defined as $\bm{n} = \bm{r}_\theta \times \bm{r}_\phi / |\bm{r}_\theta \times \bm{r}_\phi|$. The differential area is $dS = \sqrt{EG - F^2} \, d\theta \, d\phi$, and $\bm{r}_\theta = \partial \bm{r}/\partial \theta$, $\bm{r}_\phi = \partial \bm{r}/\partial \phi$, $\bm{r}_{\theta\theta} = \partial^2 \bm{r}/\partial \theta^2$, $\bm{r}_{\phi\phi} = \partial^2 \bm{r}/\partial \phi^2$, $\bm{r}_{\theta\phi} = \partial^2 \bm{r}/\partial \theta \partial \phi$ are the first- and second-order partial derivatives of $\bm{r}$ with respect to $\theta$ and $\phi$.

This approach provides an analytical framework for validating the polyhedral surfaces, enabling a comparative morphological analysis of the \textit{cores}.

\subsubsection{Dimensions and global topology of \textit{cores}}

Once the MFs have been determined using both the polyhedral and ellipsoidal surface methods, the shapefinders can be directly computed using the definitions in Equations ~\eqref{Ts} to \eqref{Fs}. The resulting values of $V$, $S$, $C$, $\chi$, $\mathcal{G}$, and the shapefinders $\mathcal{T}$, $\mathcal{B}$, $\mathcal{L}$, $\mathcal{P}$, and $\mathcal{F}$ for each DCC \textit{core}, based on polyhedral (with $s_\text{f}=1$) and ellipsoidal surface fits, are summarized in Tables~\ref{tab:core_poly} and \ref{tab:core_ell}, respectively.

The shapefinders $\mathcal{T}$, $\mathcal{B}$, and $\mathcal{L}$, which have units of length, are particularly useful for estimating the dimensions of the \textit{cores}. The smallest shapefinder, $\mathcal{T}$, represents the thickness of the \textit{cores}; the intermediate shapefinder, $\mathcal{B}$, is analogous to the breadth (width); and the largest shapefinder, $\mathcal{L}$, characterizes the length. It should be noted that $\mathcal{L}$ is not the actual physical length of the structure, but rather a morphological measure related to the integrated curvature of its surface \citep{ei2007a}. Thus, $\mathcal{L}$ can become significantly large for irregularly shaped or curved surfaces, without necessarily corresponding directly to a linear extension, as in elongated structures. The mean and median values of these dimensions are presented in Table~\ref{tab:SF}. As shown in the table, the extension in any dimension of the DCC \textit{cores} (as measured by $\mathcal{T}$, $\mathcal{B}$, and $\mathcal{L}$) does not exceed $10 \, h_{70}^{-1}\,\mathrm{Mpc}$ for either surface method. These results are consistent with those of \citet{ei2007a}, who reported similar scales for the densest central regions of superclusters constrained to high-mass fraction areas, which coincide with their \textit{cores}.

{
Although the polyhedral and ellipsoidal surface fits share the same boundary-point selection process, the mean (and median) values of $\mathcal{T}$, $\mathcal{B}$, and $\mathcal{L}$ obtained with these two methods differ significantly. These differences stem from the intrinsic nature of each approach. The polyhedral fit with $s_{\rm f}=1$ yields a more compact representation of the surface enclosing the member galaxies of the \textit{cores}, capturing (via triangulation) finer substructures and morphological details that are typical of dynamically evolving regions. Consequently, this compactness produces smaller values of $\mathcal{T}$, $\mathcal{B}$, and $\mathcal{L}$, since, by construction, surface area and volume are minimal at $s_{\rm f}=1$ (and increase toward $s_{\rm f}=0$). In contrast, the ellipsoidal fit enforces a smoother, global shape that encloses most of the \textit{core} member galaxies within an ellipsoidal boundary, which systematically returns larger values for the shapefinders.} 

\begin{table}
\caption[Mean and median values for the $\mathcal{T}$, $\mathcal{B}$, and $\mathcal{L}$ shapefinders estimated from polyhedral and ellipsoidal surface fits to the DCC \textit{cores}.]{Mean (with standard deviation) and median values for the $\mathcal{T}$, $\mathcal{B}$, and $\mathcal{L}$ shapefinders estimated from polyhedral and ellipsoidal surface fits to the DCC \textit{cores}. Median values are given as asymmetric ranges with $\Delta\text{Q}_1=\mathrm{Median}-\text{Q}_1$ and $\Delta\text{Q}_3=\text{Q}_3-\mathrm{Median}$ as lower and upper indices, where $\text{Q}_1$ and $\text{Q}_3$ are the 25th and 75th percentiles, respectively.}

\label{tab:SF}
\centering
\resizebox{\columnwidth}{!}{
\begin{tabular}{c cc cc}
\hline\hline
\multirow{2}{*}{} & \multicolumn{2}{c}{Polyhedral surfaces} & \multicolumn{2}{c}{Ellipsoidal surfaces}\\ 
 Shapefinder & Mean $\pm$ std  & Median$^{+\Delta Q_3}_{-\Delta Q_1}$ & Mean $\pm$ std  & Median$^{+\Delta Q_3}_{-\Delta Q_1}$ \\ \hline
$\mathcal{T}$ [$h_{70}^{-1}$Mpc] & $1.52 \pm 0.50$ & $1.43_{-0.24}^{+0.34}$ & $5.70 \pm 1.51$ & $5.45_{-0.92}^{+1.34}$ \\
$\mathcal{B}$ [$h_{70}^{-1}$Mpc] & $2.47 \pm 0.91$ & $2.26_{-0.43}^{+0.61}$ & $6.75 \pm 1.52$ & $6.48_{-0.83}^{+1.33}$ \\
$\mathcal{L}$ [$h_{70}^{-1}$Mpc] & $4.88 \pm 2.31$ & $4.40_{-1.19}^{+1.67}$ & $8.07 \pm 1.80$ & $7.89_{-1.06}^{+1.42}$ \\ \hline
\end{tabular}}
\end{table}

{
This dual strategy (compact polyhedral fits for structural detail and ellipsoidal fits for global extent) provides a robust morphological characterisation of the \textit{cores}: the results of the polyhedral fit can be interpreted as a lower bound on the dimensions of these structures, while those of the ellipsoidal fits serve as an upper bound. The ellipsoidal method is included for comparison because it is a conservative technique widely used in the literature. Furthermore, exploratory tests (not included in the manuscript for brevity) employing intermediate shrink factors (e.g., $0 \le s_{\rm f} < 1$) in polyhedral fits confirm that the qualitative classification, in particular the dominance of filamentary morphologies, remains consistent and does not critically depend on the specific choice of $s_{\rm f}$.}

\begin{figure} %<left> <lower> <right> <upper> 
\centering 
\includegraphics[trim={2.5cm 2.1cm 2.5cm 0.9cm},clip,width=\columnwidth]{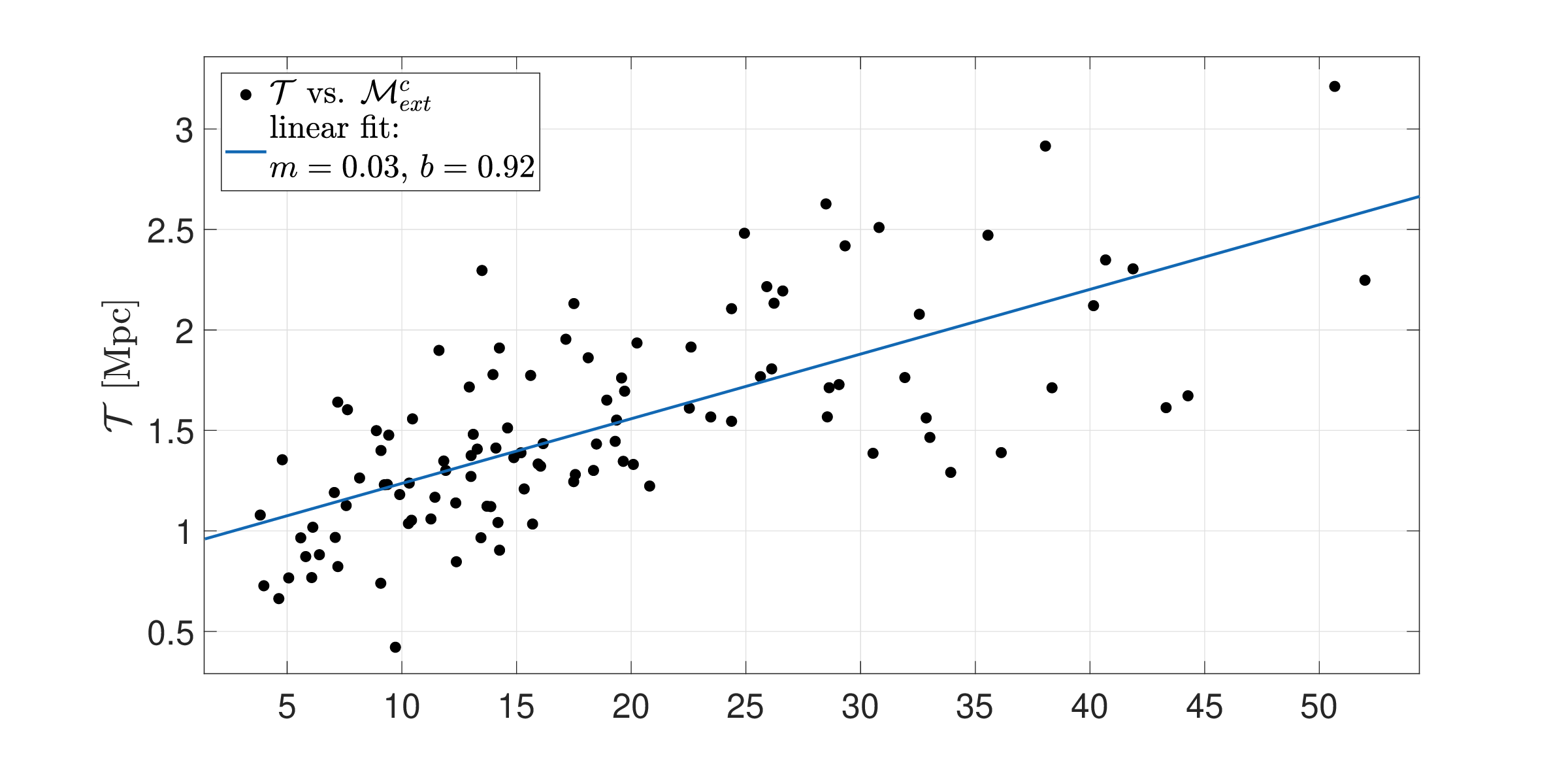} \\
\includegraphics[trim={2.5cm 2.1cm 2.5cm 1.4cm},clip,width=\columnwidth]{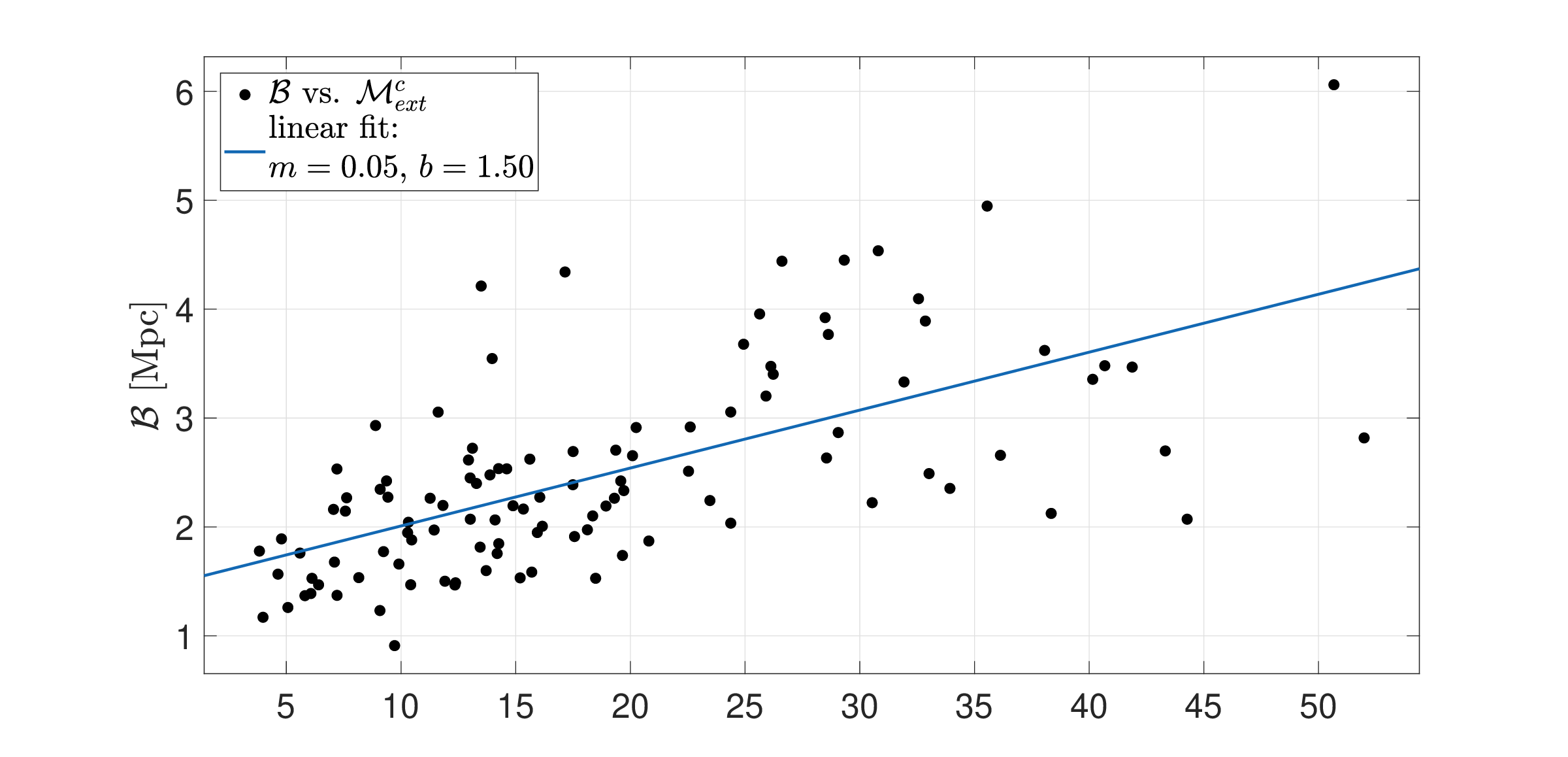} \\
\includegraphics[trim={2.5cm 0cm 2.5cm 1.4cm},clip,width=\columnwidth]{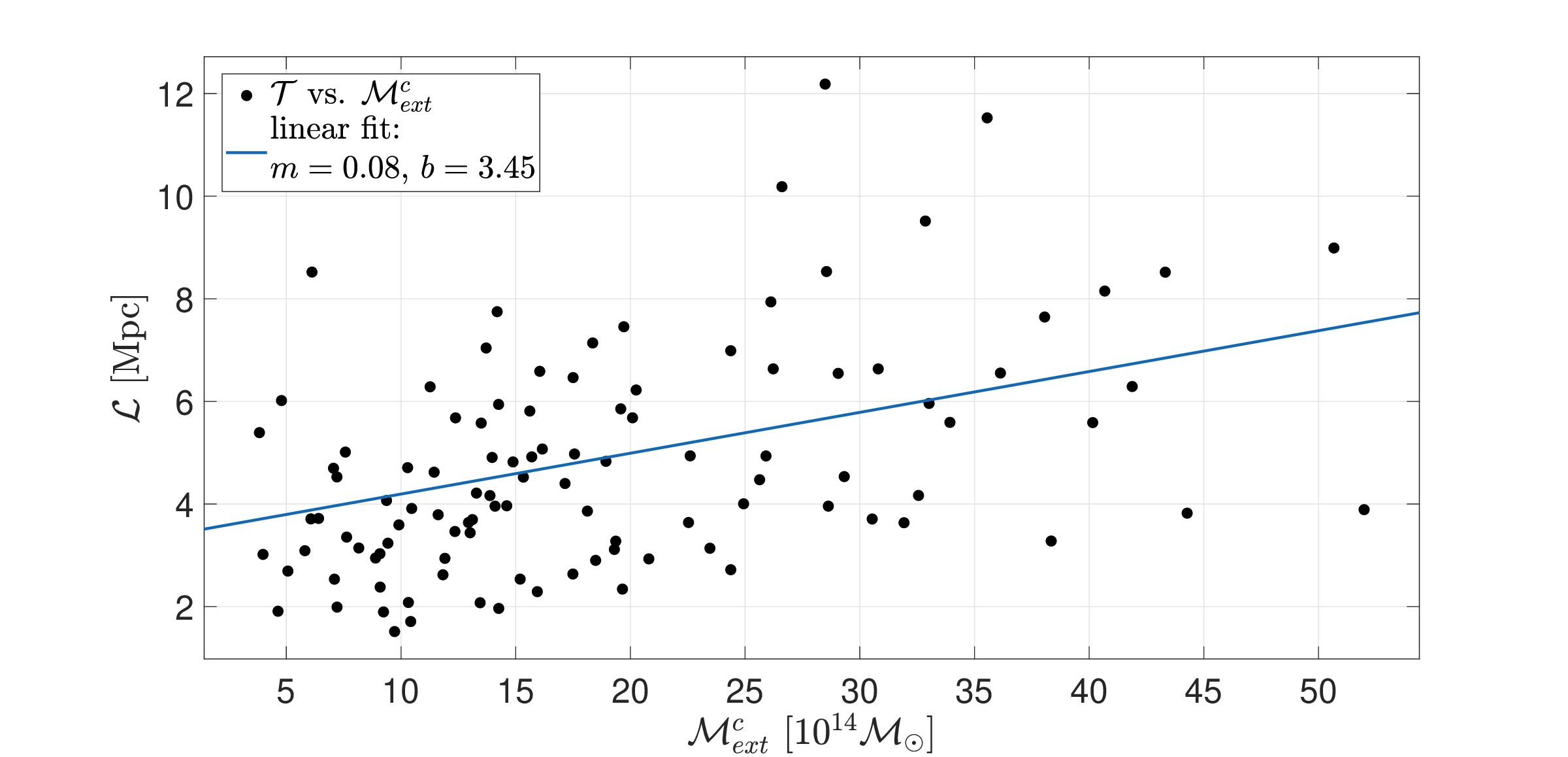} \\
 \caption[Length, breadth and thickness ($\mathcal{T}$, $\mathcal{B}$ and $\mathcal{L}$ shapefinders) versus extensive mass for DCC \textit{cores}.]{Length, breadth and thickness ($\mathcal{T}$, $\mathcal{B}$ and $\mathcal{L}$ shapefinders) versus extensive mass $\mathcal{M}_\text{ext}^c$ for DCC \textit{cores}. The three solid blue lines correspond to the best linear fit in each case. {The slope $m$ and vertical intercept $b$ of each fit are displayed in the upper-left inset of the corresponding panel}. The units of each axis must be understood in terms of $h_{70}^{-1}$.} 
 \label{f:TBL_M}
 \end{figure}

Figure~\ref{f:TBL_M} shows scatterplots of the three characteristic dimensions ($\mathcal{T}$, $\mathcal{B}$, and $\mathcal{L}$ from Table \ref{tab:core_poly}) of the DCC \textit{cores} versus their extensive mass ($\mathcal{M}_\text{ext}^c$, defined as the sum of the virial masses of the member galaxy systems of a \textit{core}, and presented in Table 5 of Paper I). Statistical correlations were evaluated using a Pearson test with a 95\% confidence level. The correlation coefficients for $\mathcal{T}$, $\mathcal{B}$, and $\mathcal{L}$ with mass were $0.70$, $0.64$, and $0.40$, respectively. These correlations are statistically significant with a significance level of $\alpha_s=0.05$. These results suggest that the dimensions of the \textit{cores} are positively correlated with their mass: more massive structures tend to exhibit greater thickness, breadth, and length. This finding is consistent with previous studies by \citet{sha2004}, which reported similar trends in superclusters.

On the other hand, a standard interpretation of the genus in the cosmology literature defines it as \citep[e.g.,][]{sha2004}:  
\begin{equation} \label{gen1}
\mathcal{G} = (\text{number of holes}) - (\text{number of isolated regions}) + 1,
\end{equation}  
where `holes' are considered complex mathematical objects that, in three-dimensional structures, typically manifest as tunnels crossing the structure from one side to the other, as in a toroidal shape. The term `isolated regions' refers to the number of disconnected parts of the fitted surface that define the structure’s boundaries. Thus, Equation~\eqref{gen2} can be rewritten as:  
\begin{equation} \label{gen2}
\mathcal{G} = N_\text{tunn} - N_\text{is,surf} + 1,
\end{equation}  
where $N_\text{tunn}$ and $N_\text{is,surf}$ represent the number of tunnels and isolated surfaces, respectively \citep[e.g.,][]{bag2020}.  

\begin{figure} %<left> <lower> <right> <upper> 
\centering 
\includegraphics[trim={2.5cm 0cm 2.5cm 1cm},clip,width=\columnwidth]{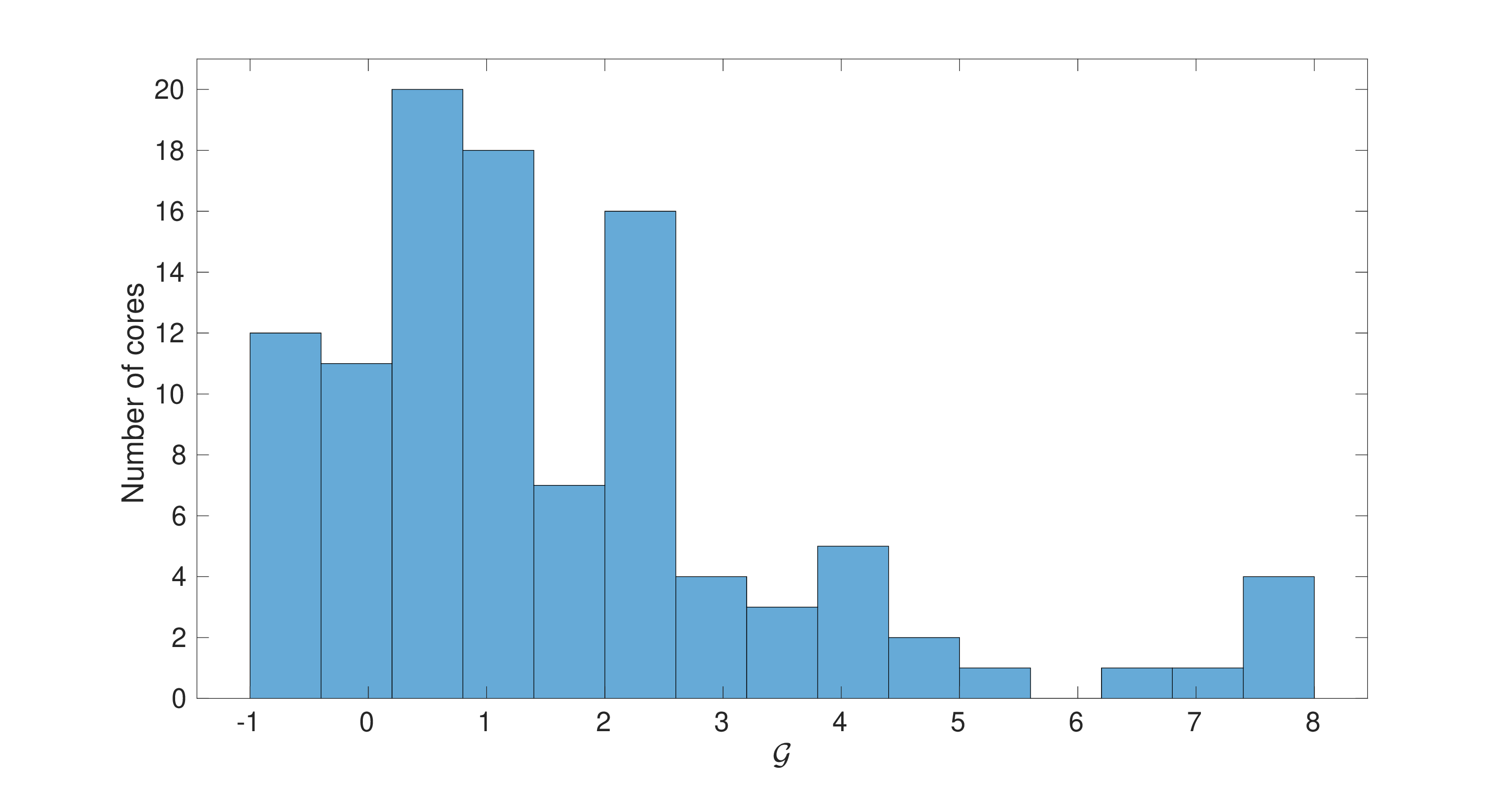} \\
\includegraphics[trim={2.5cm 0cm 2.5cm 1cm},clip,width=\columnwidth]{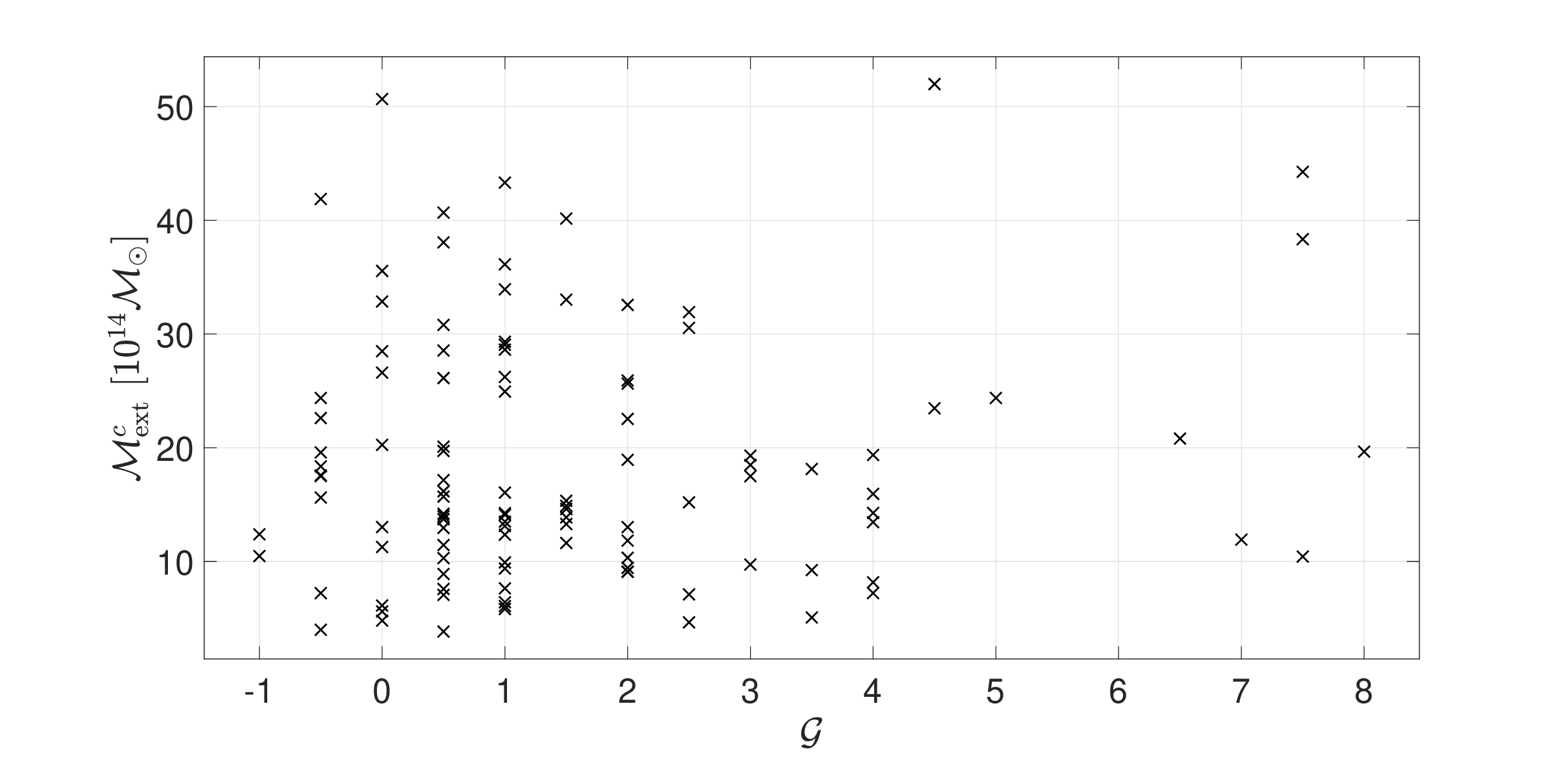} \\
 \caption[Distribution of (genus) $\mathcal{G}$ values for DCC \textit{cores}.]{\textit{Top}: Distribution of (genus) $\mathcal{G}$ values for DCC \textit{cores}. \textit{Bottom}: Distribution of $\mathcal{G}$ values versus the extensive mass $\mathcal{M}_\text{ext}^c$ of DCC \textit{cores}.} 
 \label{f:gen}
 \end{figure}

Visual inspections using basic rotation in 3D and rendering tools available in standard scientific software (e.g., MATLAB and Python libraries) revealed that the DCC \textit{cores} exhibit significant substructure patterns, frequently containing tunnels and isolated systems. The top panel of Figure~\ref{f:gen} shows the distribution of $\mathcal{G}$ values obtained from the fitted polyhedral surfaces of the DCC \textit{cores}, as presented in Table~\ref{tab:core_poly}. The wide range of $\mathcal{G}$ values (from $-1$ to $8$) reflects the diversity of complex three-dimensional shapes exhibited by these structures. Interpreting the genus as a structural feature of such surfaces is not unique and becomes non-trivial when the regions have highly intricate geometries \citep[e.g.,][]{sha2004}. 

For this analysis, the genus obtained from ellipsoidal surface fits was not considered, since ellipsoids are topologically homeomorphic to spheres, which have genus $\mathcal{G}=0$. As a result, a meaningful topological study based on this type of surface fit is not feasible.  

Various non-integer values of $\mathcal{G}$ were obtained because the genus was computed using the formula $\mathcal{G}=1-\chi/2$, where $\chi$ is the Euler characteristic. Since $\chi$ can take odd values depending on the discretisation and resolution of the surface representation, $\mathcal{G}$ is not always an integer. However, genus is fundamentally a topological invariant that describes the number of handles or holes in a surface, and it is conventionally defined for integer values. Non-integer values are difficult to interpret in this context, as they do not correspond to well-defined topological classes. Therefore, we restrict our analysis to integer values as defined in Equation~\eqref{gen2}. Under this criterion, about 10\% of the DCC \textit{cores} are topologically isomorphic to a sphere (i.e., $\mathcal{G}=0$), $\sim$18\% to a toroid (i.e., $\mathcal{G}=1$), and $\sim$10\% to a pretzel (i.e., $\mathcal{G}=2$). Higher genus values indicate increasingly complex topologies. Structures with negative genus (about 11\% of the \textit{cores}) can be interpreted as coarser configurations with multiple isolated member systems.

Additionally, the bottom panel of Figure~\ref{f:gen} shows the distribution of all genus values (both integer and non-integer) versus the extensive mass of \textit{cores}, though no significant statistical correlation can be inferred from this relationship. This suggests that the topological complexity of \textit{cores} is not directly governed by their mass, possibly due to the influence of other structural and dynamical factors.

\subsubsection{The spectrum of shapes of the \textit{cores}} \label{shape_s}

\begin{figure} %<left> <lower> <right> <upper> 
\centering 
\includegraphics[trim={1cm 0.8cm 1.5cm 1cm},clip,width=\columnwidth]{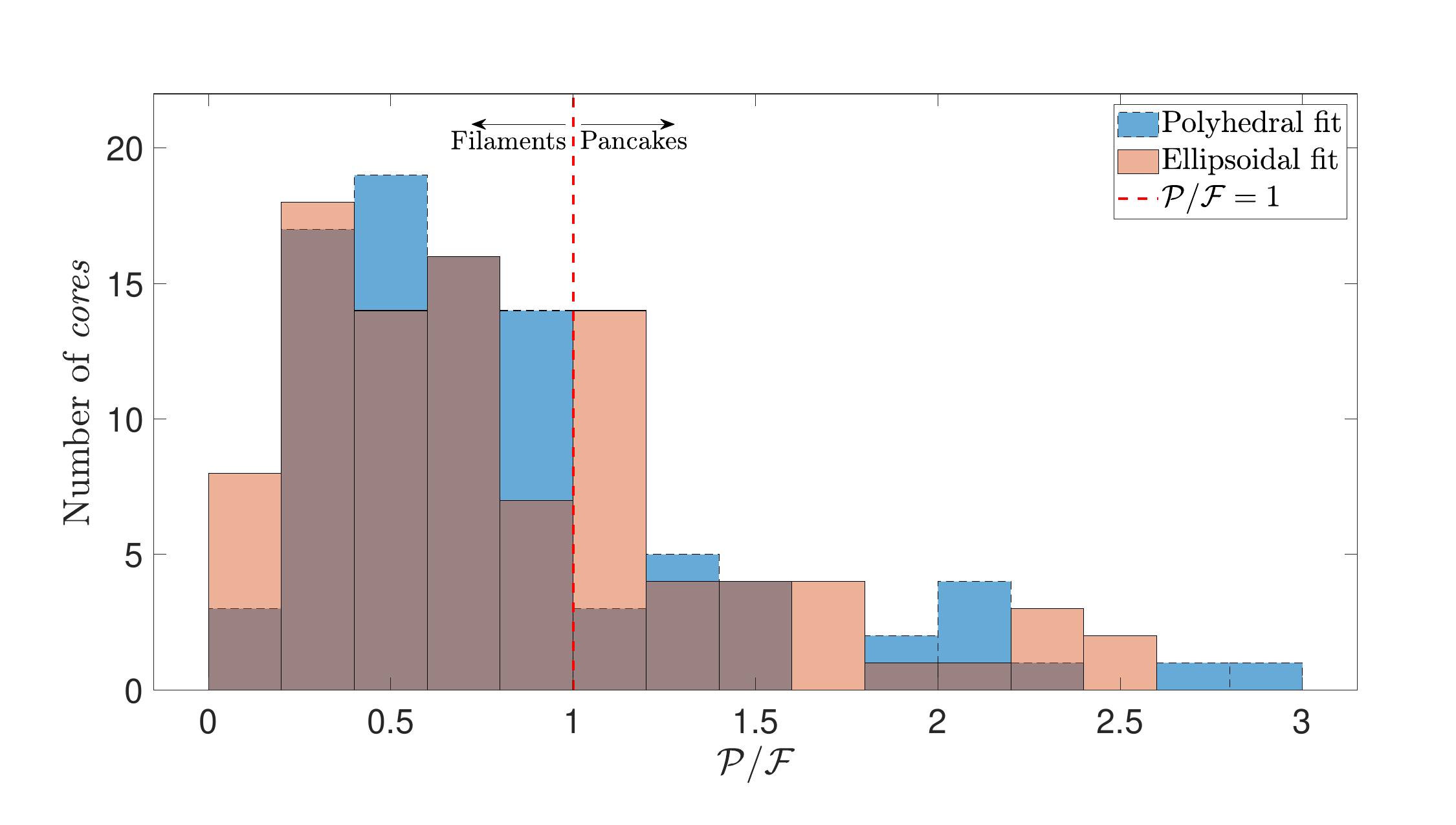} \\
 \caption[The shape spectrum of DCC \textit{cores}]{The `shape spectrum' of DCC \textit{cores}. Distribution of the shape statistic $\mathcal{P}/\mathcal{F}$ in the sample of \textit{cores} excluding outliers and using two surface fit (polyhedral and ellipsoidal) methods. The structures are classified as filaments if $0\leq \mathcal{P}/\mathcal{F}\leq 1$ or as pancakes if $\mathcal{P}/\mathcal{F}>1$.} 
 \label{f:fil_pan}
 \end{figure}

In addition to MFs and shapefinders, Tables \ref{tab:core_poly} and \ref{tab:core_ell} present the morphological parameter $\mathcal{P}/\mathcal{F}$, which enables the statistical analysis of the shapes of the \textit{cores}. The shapefinder formalism allows for classifying \textit{cores} into two basic morphological types:
\begin{enumerate}
\item Pancakes (oblate structures), with $\mathcal{P}/\mathcal{F}>1$, and
\item Filaments (prolate structures), with $0\leq\mathcal{P}/\mathcal{F}\leq1$.
\end{enumerate}
Here, ribbon-like objects were excluded, as defining the range $\mathcal{P}/\mathcal{F}\approx 1$ for this morphology is somewhat arbitrary \citep[e.g.,][]{cd2011}. The distribution of the morphological parameter $\mathcal{P}/\mathcal{F}$, referred to as the \textit{shape spectrum}, for the DCC \textit{cores} is shown in Figure \ref{f:fil_pan}. Outliers (15 data points from polyhedral fits and 9 from ellipsoidal fits) were excluded from the analysis.
The polyhedral surface fits reveal that about 75\% of the \textit{cores} can be classified as filaments, while the remaining $\sim$25\% can be identified as pancakes. The ellipsoidal fits corroborate these findings, indicating that approximately 62\% of the \textit{cores} have filamentary morphologies, while $\sim$38\% exhibit pancake-like shapes. The shape spectrum obtained from both methods confirms a statistical tendency of DCC \textit{cores} toward filamentary morphologies.

Despite the differences in the values of the characteristic dimensions (see Table \ref{tab:SF}) obtained from polyhedral and ellipsoidal surface fitting, the analyses of planarity and filamentarity remain consistent for both methods. Since these quantities are defined as combined ratios of $\mathcal{T}$, $\mathcal{B}$, and $\mathcal{L}$ (see Equations \ref{Ps} and \ref{Fs}), their stability suggests that both fitting approaches preserve the overall structural characteristics of the \textit{cores} (see Sect. \ref{shape_s}). This reinforces the robustness of our morphological classification, regardless of the specific method used to estimate absolute dimensions. Therefore, while the polyhedral fit provides a more precise morphological description, the ellipsoidal fit remains a useful comparative reference within the framework of large-scale structure studies.

Since filaments are the most common features in rich superclusters \citep[e.g.,][]{ei2007a,cau2014}, it is reasonable to expect that \textit{cores}, being internal substructures of superclusters, tend toward these morphologies. This is consistent with the hierarchical growth these structures experience, as \textit{cores} are primarily `nourished' through the filaments to which they are connected. Filaments act as massive channels for transporting matter in the Universe \citep[e.g.,][and references therein]{lib2018}, primarily along directions of maximum anisotropic gravitational attraction \citep[e.g.,][]{zel1970}, in this case toward \textit{cores}, which are the densest regions within superclusters.

On the other hand, pancake-like \textit{cores} are less common in rich superclusters but represent another class of structures that can evolve into compact, virialised objects. Pancake \textit{cores} are more prevalent in lumpy, less filamentary superclusters and are often isolated or disconnected from other parts of their host superclusters.

The predominantly anisotropic morphologies (flattened and elongated) of \textit{cores} are indicative of their current quasi-linear dynamical stage, as predicted by the Zeldovich model \citep[e.g.,][]{am2009}. As \textit{cores} evolve, their morphologies change, leading to various intermediate shapes over time. According to the Zeldovich formalism, which defines a morphological sequence linked to stages of an anisotropic gravitational collapse \citep[see, e.g.,][]{cau2014}, \textit{cores} that are mostly filaments are the largest bound structures closest to becoming virialised objects, as they only need to collapse along a single axis (the longest of the three). In contrast, pancake-shaped \textit{cores} still need to collapse along two axes before reaching virialisation. However, since they are already bound structures with significant overdensities, their collapse and future virialisation are inevitable.

\begin{figure} %<left> <lower> <right> <upper> 
\centering 
\includegraphics[trim={2.5cm 0.5cm 2.5cm 0.5cm},clip,width=\columnwidth]{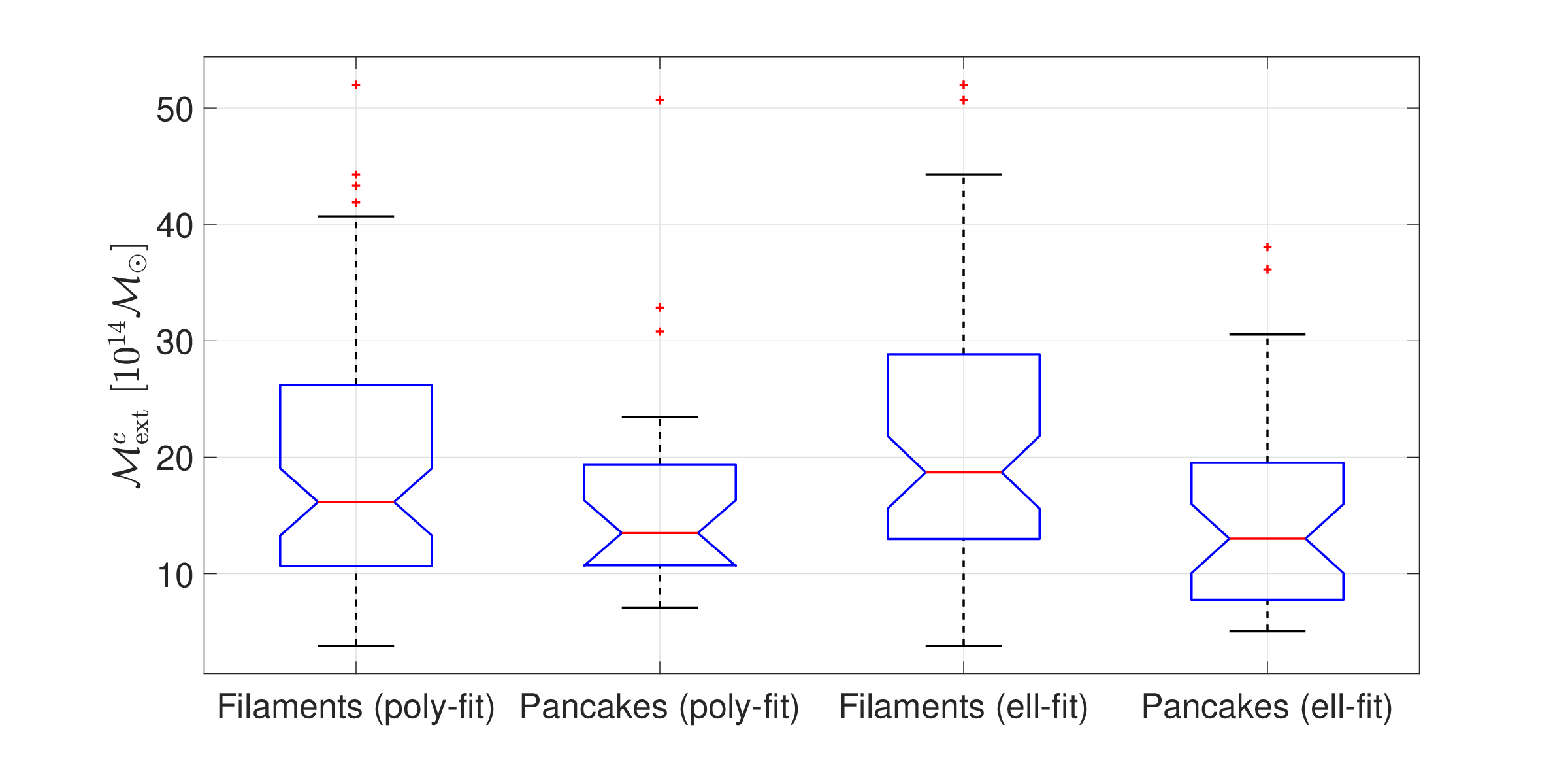} \\
 \caption[Boxplots of extensive mass for the two morphological classifications of DCC \textit{cores}: filaments and pancakes.]{Boxplots of extensive mass $\mathcal{M}_\text{ext}^c$ for the two morphological classifications of DCC \textit{cores}, filaments ($0\leq \mathcal{P}/\mathcal{F}\leq 1$) and pancakes ($\mathcal{P}/\mathcal{F}>1$), based on the polyhedral (poly-fit) and ellipsoidal (ell-fit) fits  surfaces. Filaments have a relatively greater mass than pancakes.} 
 \label{f:FP_M}
 \end{figure}

The evolution of \textit{cores} rarely occurs in isolation within the LSS. As previously noted, most \textit{cores} are dense filaments or located at filament intersections within superclusters, constantly accreting matter from their surroundings. The boxplot in Figure \ref{f:FP_M} shows that filamentary \textit{cores} ($0\leq\mathcal{P}/\mathcal{F}\leq 1$) are relatively more massive than pancakes ($\mathcal{P}/\mathcal{F}>1$). The mass distribution of filaments extends toward higher values, while pancake masses are distributed within a narrower, lower-mass range. Pancake-like \textit{cores} are often surrounded by less dense regions, as evidenced by density contrast analyses and connectivity studies between \textit{cores} and their host superclusters. As \textit{core} richness increases with mass, these results align with \citet{cd2011}, who found that filaments in superclusters tend to be richer than pancakes on average.

Finally, we estimate the fraction of \textit{cores} with approximately spherical shapes (i.e., $\mathcal{P} \approx \mathcal{F} \approx 0$). Allowing a tolerance of up to 0.05 in both $\mathcal{P}$ (planarity) and $\mathcal{F}$ (filamentarity), only $\sim$8\% of the DCC \textit{cores} are approximately spherical based on ellipsoidal fits. None of these \textit{cores} fall within this range when using polyhedral surface fits. Decreasing the tolerance further reduces the fraction of \textit{cores} with comparable axes ($a\approx b\approx c$). The virtual absence of spherical objects underscores the relative youth of the \textit{cores}, which may reflect the fact that the primordial density field did not contain spherical overdense regions \citep[e.g.,][]{bar1986,kra2012}, and that the early stages of contraction and gravitational collapse occur in strongly flattened and elongated geometries \citep[e.g.,][]{zel1970,am2009,cau2014}.

\section{Entropy Analysis}\label{entropy}

According to the standard model of structure formation, gravitational collapse occurs anisotropically, progressing from flattened and elongated configurations to a more compact triaxial structure in a virialised state \citep[e.g.,][]{sha2004,cau2014}. This collapse along different axes leads to a continuous redistribution of mass, shaping the evolving density profiles of these structures \citep[e.g.,][]{am2009}. In the case of \textit{cores} --- highly overdense regions within superclusters --- their internal structure is primarily shaped by the spatial distribution of their member galaxy systems, as traced in optical observations. As \textit{cores} dynamically evolve, it is expected that smaller groups accrete onto richer clusters, which in turn merge, gradually dissolving substructures and leading to a more homogeneous (spatial and velocity) galaxy distributions similar to those in relaxed massive clusters. The most massive cluster (MMC) within each \textit{core} could eventually become the main gravitational centre, around which the virialised structure of the \textit{core} may form.

The evolution of structures in the Universe is shaped by a complex combination of fundamental physical laws and stochastic processes, often challenging our understanding \citep[e.g.,][]{lyn1967,sas1980,pd1990,bt2008}. This complexity increases further when considering environmental interactions, as matter and energy exchanges can significantly influence their dynamical evolution. One of the most significant principles influencing the evolution of any physical system is the second law of thermodynamics, which asserts that the entropy of an isolated system increases over time, reaching its maximum at equilibrium --- the most evolved state possible for a system of given mass and energy. In this context, entropy directly correlates with the degree of dynamical evolution of a physical system, making it a relevant parameter for characterizing the dynamical state of the \textit{cores} (or any other galaxy system or structure).

Entropy measures the randomness and absence of macroscopic motions or special configurations such as substructures within a system \citep[e.g.,][]{ll1980}. In this sense, it indicates how close a system is to dynamical relaxation (equilibrium). However, entropy remains a state function related to the thermodynamic properties of macroscopic systems, which are still challenging to describe and interpret in systems dominated by gravity due to their unique behaviors \citep[see][for further discussion]{zu2024b}. 

\subsection{The $H_Z$-entropy estimator}
To quantitatively characterise the evolutionary state of galaxy systems within an entropy-increasing framework, a continuous entropy estimator has been introduced \citep[see][]{zu2024b}, defined as:  

{
\begin{equation}\label{Hz0}
H_Z\equiv
\ln{\left( \frac{\mathcal{M}_\text{vir}}{\frac{4}{3}\pi R_\text{vir}^3} \frac{1}{\rho_0} \right)} +
\frac{\beta}{2}\ln{\left( \frac{\beta}{2} \frac{\sigma_v^2}{\sigma_{v_0}^2} \right)},
\end{equation}
}

where $\mathcal{M}_\text{vir}$, $R_\text{vir}$, and $\sigma_v$ denote the virial mass, virial radius, and line-of-sight galaxy velocity dispersion of a system, respectively. The parameter $\beta$ accounts for the velocity anisotropy of galaxies \citep[$\beta=3$ for Maxwellian distributions, e.g.,][]{tu2015}. {The constants $\rho_0=(10^{14}\mathcal{M}_\odot/\text{Mpc}^3)h_{70}^2$ and $\sigma_{v_0}= 1 \,\ \text{km s}^{-1}$ are fiducial reference values for density and velocity dispersion used to render the arguments of the logarithms dimensionless}. The $H_Z$ estimator is a physically motivated, dimensionless quantity that traces the net change in specific entropy ($\Delta s$) experienced by a system from its formation to the present epoch. Since it is defined in terms of (optical) observational parameters, it can be readily computed once a sufficient number of member galaxies are identified.  

The $H_Z$ estimator was first tested on a sample of 70 well-sampled galaxy clusters in the Local Universe \citep[the \textit{Top70} cluster sample\footnote{See \url{www.astro.ugto.mx/recursos/HP_SCls/Top70.html}.},][]{ca2023}. The test results suggest a strong correlation between $H_Z$ and the dynamical state of clusters \citep{zu2024b}. Specifically, $H_Z$ correlates with the level of gravitational assembly, yielding lower values for dynamically younger, substructured clusters and higher values for more relaxed, evolved systems. Furthermore, $H_Z$ exhibits a robust correlation with other continuous dynamical indicators derived from both optical and X-ray observations.

\subsection{Entropy of \textit{cores}}

The use of the $H_Z$ estimator can be extended to the study of structures larger than clusters due to self-similarity. Gravity-dominated systems are expected to follow a comparable evolutionary path on a global scale \citep[e.g., simulation results indicate that supercluster-scale structures tend to evolve into very rich cluster-like systems,][]{am2009}. Based on its successful application to characterise the dynamical state of galaxy clusters \citep{zu2024b}, we propose here to extend the use of the $H_Z$ estimator to larger structures, such as superclusters and their \textit{cores}. To this end, we generalise the $H_Z$ estimator as follows:  

{
\begin{equation}\label{Hz}
H_Z=\ln{\left( \frac{\bar{\rho}}{\rho_0} \right)} +
 \frac{\beta}{2}\ln{\left( \frac{\beta}{2} \frac{\sigma_v^2}{\sigma_{v_0}^2} \right)},
\end{equation}
}

where $\bar{\rho}$ and $\sigma_v$ represent the average mass density (e.g., $\bar{\rho}=\mathcal{M}_\text{ext}/V$) and the line-of-sight galaxy velocity dispersion of the structure under study, respectively. This generalisation avoids using virial parameters when estimating entropy for non-relaxed systems.

Using the average densities and galaxy velocity dispersions estimated for the \textit{cores} (see Table 5 of Paper I and Table \ref{tab:core_prop}) and superclusters (see Table 3 of Paper I and Table \ref{tab:SCprop}), we computed the $H_Z$-entropy for each of these structures. The $H_Z$ values obtained for the complete MSCC-supercluster sample are listed in column 6 of Table \ref{tab:SCprop}, while the $H_Z$ values for DCC \textit{cores} are shown in column 12 of Table \ref{tab:core_prop}. 

To compare evolutionary states of galaxy systems and structures at different scales, we took the $H_Z$-entropy values for the \textit{Top70} cluster sample from Table 2 of \citet{zu2024b}. These clusters, among the best-sampled galaxy systems in the nearby Universe, belong to MSCC superclusters and cover a broad range from poor to rich systems (with ICM temperatures from 1 to 12 keV). 

\begin{figure} %<left> <lower> <right> <upper> 
\centering 
\includegraphics[trim={2.5cm 0.5cm 3cm 1cm},clip,width=\columnwidth]{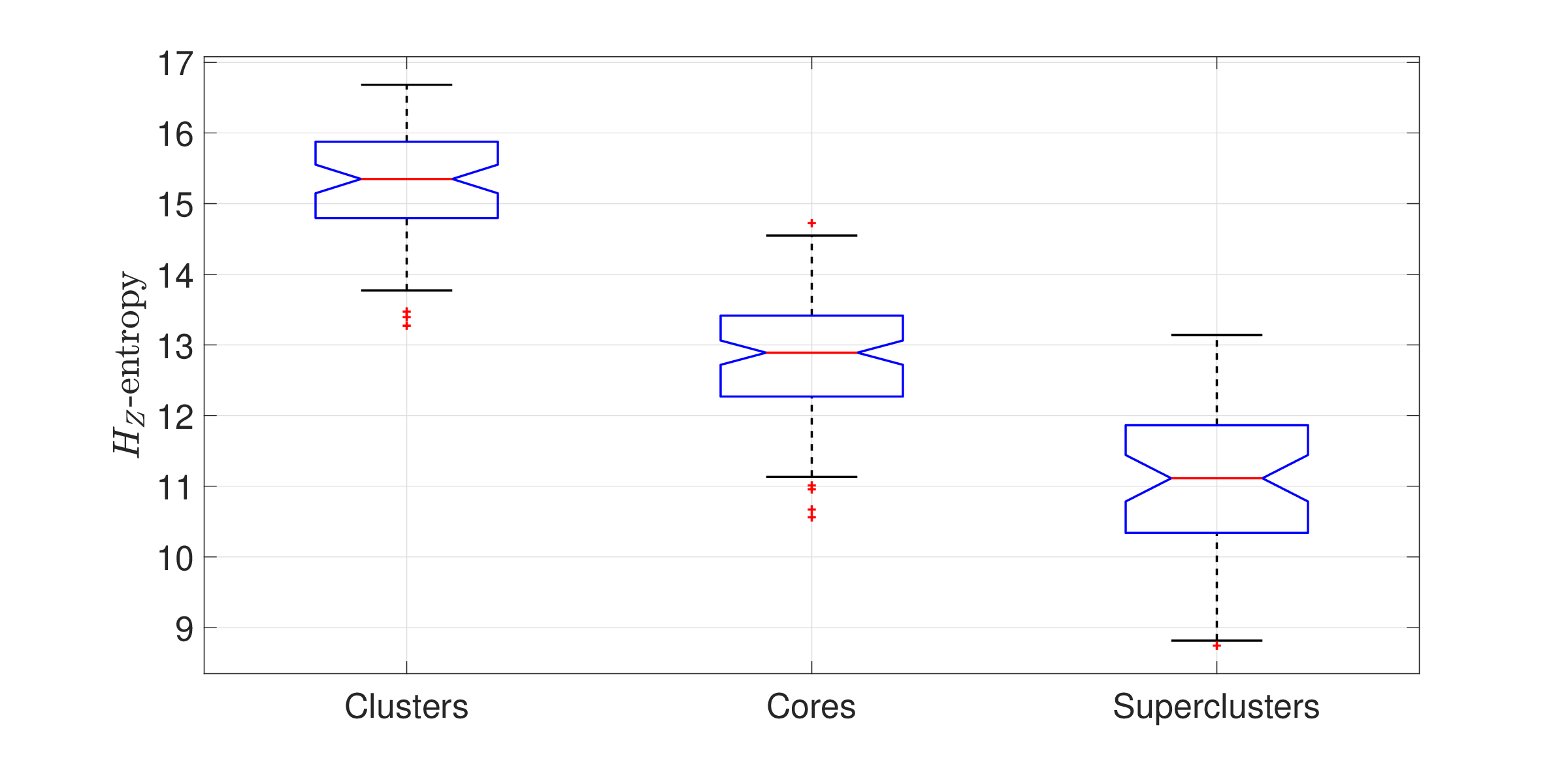} \\
\includegraphics[trim={2.5cm 0.5cm 3cm 1cm},clip,width=\columnwidth]{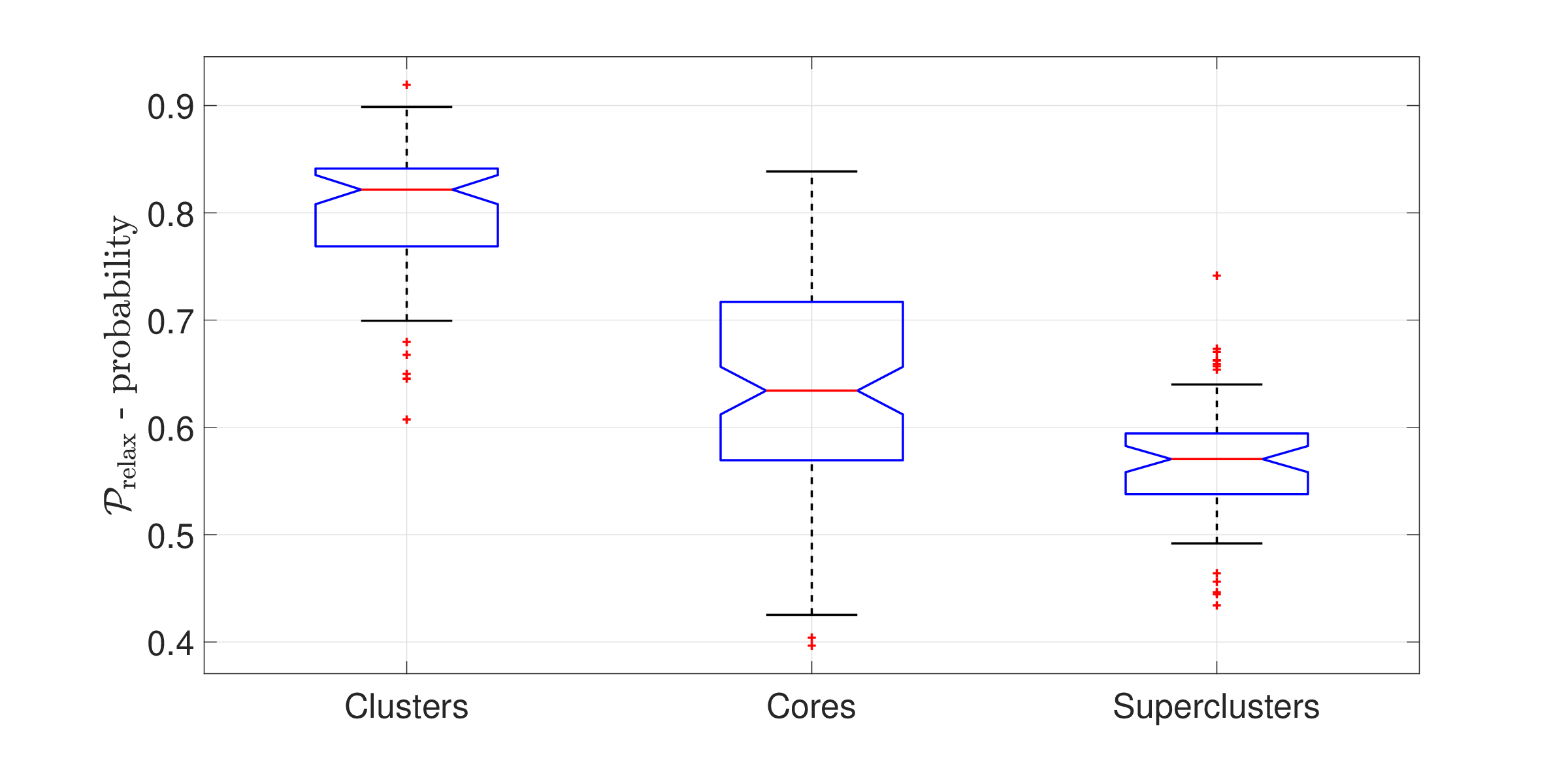} \\
 \caption[Distributions of $H_Z$ entropy values for clusters, \textit{cores} and superclusters.]{Distributions of $H_Z$-entropy (top panel) and probability $\mathcal{P}_\mathrm{relax}$ (bottom panel) values for clusters (the \textit{Top70} sample), \textit{cores} (from DCC) and superclusters (from MSCC).} 
 \label{f:Hz}
 \end{figure}

The top panel of Figure \ref{f:Hz} displays boxplots of $H_Z$-entropy distributions for clusters, \textit{cores}, and superclusters. It is evident that clusters exhibit higher entropies (median value of $15.4$), consistent with systems near virial equilibrium. In contrast, superclusters show the lowest entropies (median value of $11.1$), indicating their less advanced evolutionary state among the studied galaxy structures. Meanwhile, \textit{cores} present intermediate entropy values (median value of $12.9$), supporting the hypothesis that these regions are dynamically more evolved than their rich host superclusters.

 \subsection{Relaxation probability of \textit{cores}}
To complement the entropy-based approach, we evaluate the relaxation probability parameter, $\mathcal{P}_{\text{relax}}$, as defined in \citet{zu2024b}. This parameter provides an independent method for assessing the dynamical state of galaxy structures based on the distribution of their member galaxies in phase space. virialised \textit{cores} are expected to exhibit spatial and velocity distributions similar to those of relaxed galaxy clusters. These distributions are typically modeled using profiles derived from equilibrium assumptions, such as the King profile for radial distributions, uniform azimuthal distributions, and Gaussian velocity distributions \citep[e.g.,][]{sas1984,sa1986,ad1998,sam2014}.

The $\mathcal{P}_{\text{relax}}$ parameter is determined by comparing the observed probability density functions (PDFs) of galaxies within a \textit{core} to equilibrium models. This approach mirrors that used in \citet{zu2024b} for galaxy clusters. Briefly, the raw observational coordinates of member galaxies, namely the triples $(\text{RA}, \text{Dec}, z)$ of right ascension, declination, and redshift, are distributed within a solid angle that can be approximated by a cylinder with a circular base in the plane of the sky and depth along the line of sight. Within this cylinder, each galaxy’s position can be expressed as $(r, \theta, z)$, where $r$ represents its projected distance from the \textit{core} centroid, $\theta$ its azimuthal angle relative to the local north direction in the projected sky distribution, and $z$ its redshift, serving as a proxy for radial velocity. Assuming statistical independence\footnote{This assumption is supported by the low fraction of galaxies exhibiting a strong correlation between $r$ and $z$, which carry negligible statistical weight and show no significant correlations between $r$, $z$, and $\theta$ \citep[see][]{zu2024b}.}, the galaxy distribution in each \textit{core} can be described by an empirical joint PDF, $\bar{f}_{r\theta z} = \bar{f}_r(r)\bar{f}_\theta(\theta)\bar{f}_z(z)$. Here, $\bar{f}_r(r)$, $\bar{f}_\theta(\theta)$, and $\bar{f}_z(z)$ represent the observed PDFs for the radial-$r$, azimuthal-$\theta$, and redshift-$z$ variables, respectively.

To estimate these observed PDFs, we apply a smoothing technique using a kernel density estimator over normalized galaxy counts within bins of width $\Delta r = 0.35\, h_{70}^{-1}$ Mpc, $\Delta \theta = 12^\circ$, and $c\Delta z = 250$ km s$^{-1}$ for the $r$, $\theta$, and $z$ variables, respectively. These bin widths were chosen from a range of test values to optimise the fit of the distribution functions during histogram smoothing. A standard Gaussian smoothing kernel is applied using the same bin widths over the intervals $[0, R_h]$, $[0, 360^\circ]$, and $[z_\mathrm{min}, z_\mathrm{max}]$ for $\bar{f}_r$, $\bar{f}_\theta$, and $\bar{f}_z$, respectively. Here, $z_\mathrm{min}$ and $z_\mathrm{max}$ are the minimum and maximum redshifts of the galaxies in the structure, and 
\begin{equation}\label{Rh} 
R_h\equiv \frac{2N(N-1)}{\sum_{i\neq j}R_{ij}^{-1}},
\end{equation}
represents the harmonic radius of the structure calculated from the projected distances $R_{ij}$ (in Mpc) between its $N$ sampled galaxies (see column 10 of Table \ref{tab:core_prop}). Since $\bar{f}_r$, $\bar{f}_\theta$, and $\bar{f}_z$ depend on the dynamical state of the structure, they describe the current distributions of its member galaxies.

To define the equivalent relaxed PDFs for each variable in each \textit{core} (or supercluster), we adopt a simple reference equilibrium model. virialised \textit{cores} are assumed to exhibit spherical galaxy distributions with a homogeneous core-halo configuration and isotropic velocities, lacking net angular momentum. For simplicity, we use the King-type radial, continuous uniform, and normal distributions ${f}_r^\mathrm{eq}(r)$, ${f}_\theta^\mathrm{eq}(\theta)$, and ${f}_z^\mathrm{eq}(z)$ from \citet{zu2024b} as equilibrium models for the $r$, $\theta$, and $z$ variables, respectively. The similarity between observed and relaxed PDFs is quantified using the Hellinger distance \citep[$0 \leq H(\bar{f}_k, f_k^\mathrm{eq}) \leq 1$;][]{he1909}, where lower Hellinger distances indicate distributions closer to equilibrium, corresponding to higher relaxation probabilities since $\mathcal{P}_\mathrm{relax} \equiv 1 - H$.

The $\mathcal{P}_\mathrm{relax}$ values computed for DCC \textit{cores} are shown in column 13 of Table \ref{tab:core_prop}. Additionally, the bottom panel of Figure \ref{f:Hz} illustrates the $\mathcal{P}_\mathrm{relax}$ distributions for the \textit{Top70} clusters, DCC \textit{cores}, and MSCC superclusters. For superclusters, these values were estimated similarly (see column 7 of Table \ref{tab:SCprop}), while the \textit{Top70} cluster values are from \citet{zu2024b}. Consistent with the $H_Z$-entropy analysis, the relaxation probability results also support the hypothesis that \textit{cores}, with an average $\left\langle \mathcal{P}_\mathrm{relax} \right\rangle = 0.64$, represent galaxy structures in an intermediate evolutionary state. Clusters, being the most relaxed structures, exhibit $\left\langle \mathcal{P}_\mathrm{relax} \right\rangle = 0.83$, while superclusters, the least relaxed, have $\left\langle \mathcal{P}_\mathrm{relax} \right\rangle = 0.56$.

\section{Estimations of \textit{core} virial masses} 

To quantify the mass content of \textit{cores}, two complementary approaches were employed. First, the extensive mass $\mathcal{M}_\text{ext}^c$ (presented in Table 5 of Paper I) that provide a lower limit for the total mass of \textit{cores}. This is because $\mathcal{M}_\text{ext}^c$ excludes contributions from the dispersed component\footnote{The dispersed component refers to those galaxies that, while located within the \textit{cores}, are not members of galaxy systems.} and additional dark matter potentially present at the scale of these structures beyond that contained in clusters. 

\begin{figure*}
\centering
\includegraphics[trim={10.75cm 0cm 9.75cm 0cm},clip,scale=0.3]{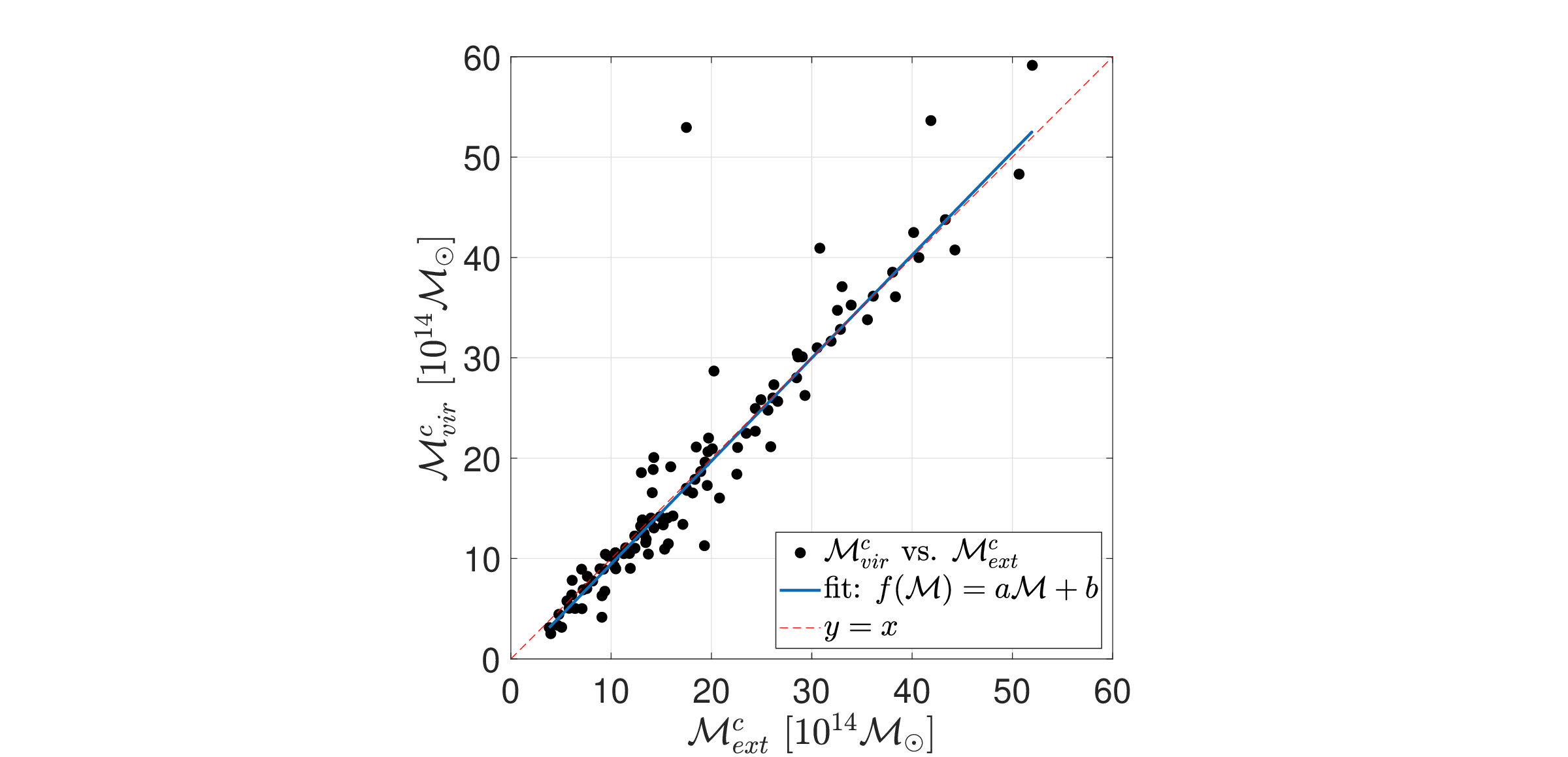} % width=0.9\textwidth, scale=0.3
\includegraphics[trim={10cm 0cm 9.9cm 0cm},clip,scale=0.3]{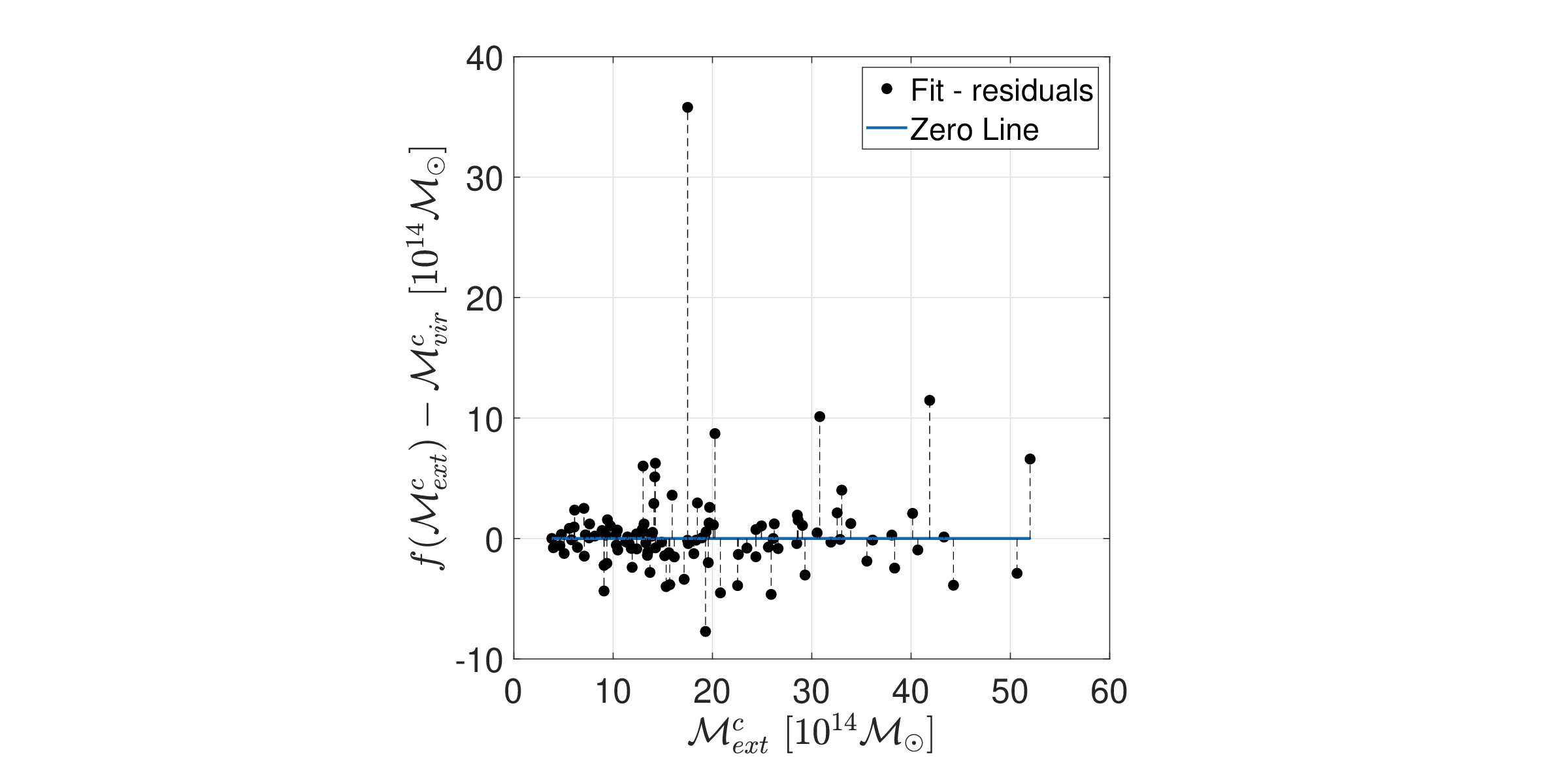}
\caption[]{Relationship between the extensive masses $\mathcal{M}_\text{ext}^c$ and the virial masses $\mathcal{M}_\text{vir}^c$ (in units of $h_{70}^{-1}$) of the DCC \textit{cores}. \textit{Left}: $\mathcal{M}_\text{vir}^c$ \textit{vs.} $\mathcal{M}_\text{ext}^c$ plot. The solid blue line represents the best linear fit $f(\mathcal{M})=a\mathcal{M}+b$ to the data, with $a=1.027$, $b=-0.917$, and a goodness of fit $R_\text{square}=0.995$. \textit{Right}: residual plot. The length of the vertical dashed lines expresses the distance between the data and the fit (zero) line.}
\label{f:vir_ext}
\end{figure*}

The second approach estimates \textit{core} masses using the virial mass estimator for clusters \citep[e.g.,][]{bi2006}:
\begin{equation}
\mathcal{M}_\text{vir}^c = \frac{\beta \pi}{2G}\sigma_v^2R_h.
\end{equation}
These calculations were based on the harmonic radii $R_h$ and line-of-sight velocity dispersions $\sigma_v$ of the galaxy distributions within the \textit{cores} (see columns 5 and 10 of Table \ref{tab:core_prop}), assuming weak anisotropy $\beta = 2.5$. The resulting $\mathcal{M}_\text{vir}^c$ values, shown in column 11 of Table \ref{tab:core_prop}, provide an estimate of the total dynamical mass. However, given that \textit{cores} are not fully virialised structures, the computed $\mathcal{M}_\text{vir}^c$ may deviate from the true values, particularly in the presence of substructures \citep[e.g.,][]{bi2006}. The most reliable estimates arise when \textit{cores} approach dynamical relaxation.

Figure \ref{f:vir_ext} illustrates the relationship between $\mathcal{M}_\text{vir}^c$ and $\mathcal{M}_\text{ext}^c$ for the DCC \textit{cores}. The correlation between these parameters is quasi-linear, with a Pearson correlation coefficient of 0.93. This trend suggests that most of the matter within \textit{cores} resides in their member galaxy systems. The dynamical evolution of \textit{cores} appears to be driven by the gravitational influence of these systems, which act as attractors, accreting surrounding matter such as individual galaxies, gas, and dark matter. This mechanism explains why the dispersed component of galaxies and external dark matter contribute minimally to the total dynamical mass. Consequently, the relationship $\mathcal{M}_\text{ext}^c \approx \mathcal{M}_\text{vir}^c$ generally holds, indicating that, before a single virialised structure is formed, all the matter in the \textit{cores} falls into the potential wells of their member systems, which dominate the macroscopic dynamics towards equilibrium.

One notable outlier, corresponding to DCC 019 (a \textit{core} of MSCC 222), deviates significantly from the linear trend. Unlike other structures such as DCC 016 and DCC 077, which exhibit pairwise gravitational binding among some member systems, DCC 019 appears unbound under any of the analyses conducted in Paper I. Its virial mass is significantly overestimated due to this lack of cohesion. Nonetheless, DCC 019 was included in the DCC catalogue owing to its high density contrast ($\mathcal{R} = 46.6$, $\Delta_\text{cr} = 14.1$), suggesting a high probability of eventual binding and virialisation.

\section{Discussion and conclusions}
The results presented in this study provide a comprehensive exploration
of the physical and dynamical properties of \textit{cores} within the
large-scale structure of the Universe. By analyzing their --- projected
and velocity --- galaxy distributions, morphology, entropy, and mass
estimates, we have deepened our understanding of these intermediate
structures and their role in the cosmic web. Below, we summarise and
discuss the key findings with respect to these aspects.

The study of the galaxy distributions and velocity dispersions within \textit{cores} shows that these structures are still far from being isotropic or dynamically relaxed. The radial velocity distributions indicate a mix of bound and infalling galaxies, supporting the view that \textit{cores} are transitional regions between virialised galaxy clusters and unbound superclusters. These dynamical states underscore the importance of the local gravitational potential and environmental effects in shaping galaxy orbits within \textit{cores}.

The morphological analysis of \textit{cores} reveals a diversity in their shapes and internal configurations, ranging from irregular and flattened structures to elongated and nearly spherical distributions (less frequent). Notably, a significant proportion of \textit{cores} exhibit filamentary morphologies, consistent with theoretical predictions from the Zeldovich approximation, which describes the anisotropic collapse of structures along preferred axes in the cosmic web. These filamentary configurations suggest that many \textit{cores} are in an intermediate evolutionary stage, where matter accretion and interactions along the connected filaments dominate their dynamics. The influence of the large-scale environment is further evident in the presence of substructures, which reflect recent mergers, accretion processes or ongoing dynamical evolution. These findings underscore the role of morphology as a diagnostic tool for understanding both the relaxation state and the formation history of \textit{cores}.

{
We note, however, that part of the observed tendency toward filamentary morphologies could, in principle, arise even in randomly distributed galaxy samples when analysed with the same polyhedral and ellipsoidal fitting procedures. Although a quantitative assessment of this effect lies beyond the scope of the present work, such a comparison would provide a useful benchmark for future studies aimed at disentangling intrinsic morphological features from those induced by statistical fluctuations or projection effects.}

The entropy analysis reveals a dynamical evolution of cosmic structures, with entropy values increasing from superclusters to \textit{cores} and galaxy clusters. This evolutionary trend, as reflected in the distributions of our entropy estimator $H_Z$, aligns with the hierarchical formation model, in which structures evolve anisotropically and gradually move toward dynamical relaxation. Clusters exhibit the highest median entropy --- and relaxation probability --- consistent with their more advanced evolutionary stage and near-virialised states. In contrast, superclusters, still in early collapse stages and characterized by strong substructure and anisotropy, display the lowest median entropy values. The intermediate entropy values observed for \textit{cores} suggest that they occupy an evolutionary stage between their less-evolved supercluster hosts and highly evolved galaxy clusters. This interpretation is consistent with the fact that \textit{cores}, as overdense regions within superclusters, are dynamically evolving structures where member galaxy systems (clusters and groups) undergo continuous merging and accretion of smaller ones, gradually dissolving substructures. Moreover, the generalized $H_Z$ estimator highlights the importance of average density and velocity dispersion in quantifying the entropy of non-virialised systems, offering a practical and robust tool for exploring the dynamical states of large-scale structures. 

Mass estimates for \textit{cores} were obtained using both extensive mass calculations, based on the sum of the virial masses of member galaxy systems, and virial mass estimates derived from the velocity dispersion and projected radii of galaxies. The strong correlation between these mass estimates ($\mathcal{M}_\text{ext}^c \approx \mathcal{M}_\text{vir}^c$) suggests that most of the matter in \textit{cores} is contained within their galaxy systems, with limited contributions from dispersed components or additional dark matter. This result supports the idea that \textit{cores} are dominated by their constituent systems, which drive the local dynamics.

Taken together, these findings provide a unified picture of \textit{cores} as dynamically evolving structures that bridge the gap between smaller, virialised systems and the larger, globally unbound supercluster environment. The interplay between gravitational forces, matter accretion, and dynamical relaxation governs their evolution, making them key sites for studying the hierarchical assembly of cosmic structures.

The main conclusions of this work are:
\begin{itemize}
    % Distribuciones espaciales y de velocidad de galaxias
    \item More than half of the \textit{cores} exhibit projected galaxy distributions consistent with the King density profile, indicating a tendency to evolve towards core-halo structures. This highlights their intermediate evolutionary status, as they begin to resemble the density profiles of more relaxed clusters.    
    
    \item Although \textit{cores} are gravitationally bound structures, analysis of the spatial and velocity distribution of their member galaxies reveals that these structures are not yet dynamically relaxed. Only about 30\% of the studied \textit{cores} show line-of-sight velocity distributions consistent with an underlying normal distribution, while the remaining $\sim$70\% do not. Since a relaxed Maxwellian distribution of galaxy velocities in three dimensions requires each velocity component to follow a normal distribution, the lack of Gaussianity in the radial velocities suggests that \textit{cores} have not yet reached dynamical equilibrium, as expected. 
    
    \item The velocity dispersions of \textit{cores} are systematically lower than those of rich galaxy clusters, even though they have comparable or greater masses. This reflects the fact that \textit{cores} are not virialised systems, so they do not exhibit a direct mass-velocity relationship like relaxed clusters. 

    % Morfología
    
\item Morphological studies reveal that \textit{cores} are mainly filamentary structures which, according to the Zeldovich approximation for large-scale structure growth, are currently in a quasi-linear dynamical stage of evolution. Given their high density contrasts, \textit{cores} have a high probability of undergoing gravitational collapse along their remaining uncollapsed dimension (i.e., the last axis of contraction), eventually transforming into virialised structures. 
    
    \item The morphology of \textit{cores} exhibits significant diversity, reflecting their dynamical states and formation histories. Many \textit{cores} display anisotropic accretion patterns and substructures, which suggest that they are still in the process of consolidating mass and transitioning toward more relaxed configurations.

    % Entropía
           \item The \textit{cores} are in a transition stage between structures in the linear evolutionary phase and fully developed virialised objects in the nonlinear stage. These results are supported by the entropy and relaxation probability analyses, which confirm the hypothesis that the \textit{cores} are structures in an evolutionary stage intermediate between superclusters and relaxed rich clusters.

       \item This work emphasises the relevance of entropy as a diagnostic tool for understanding the dynamical state and evolutionary history of cosmic structures. It provides a quantitative framework to compare galaxy systems and larger structures across different scales and stages of evolution.
      
    % Masa
    \item Mass estimates indicate that most of the matter within \textit{cores} is concentrated in their member galaxy systems. This highlights the efficiency of accretion processes in consolidating matter into gravitationally bound structures capable of surviving cosmic expansion.
    \item Within most \textit{cores}, member galaxy systems may have cleaned up their surroundings, accreting the surrounding matter (e.g, individual galaxies, gas and dark matter). This would explain why neither the dispersed component of galaxies nor the dark matter outside the clusters contribute significantly to the total dynamical mass, so that in general $\mathcal{M}_\text{ext}^c\approx \mathcal{M}_\text{vir}^c$ holds.
\end{itemize}

This study contributes to the broader understanding of \textit{cores} within the cosmic web and lays the groundwork for future investigations into their role in galaxy evolution and large-scale structure formation.

%tables:

%%%%%%%%%%%%%%%%%%%%%%%%%%%%%%%%%%%%%%%%%%%%%%%%%%%%%%%%%%%%%%%%
\begin{table*}
\begin{center}
\caption[]{Summary of the spatial, velocity, and dynamical properties of DCC \textit{cores}.}  \label{tab:core_prop}
\resizebox{18cm}{!}{
\begin{tabular}{c ccc ccc rr rrrr crcc}
\hline
\hline
  \multicolumn{1}{c}{ID} &
  \multicolumn{3}{c}{$h_\text{test}$} &
  \multicolumn{3}{c}{$p-$values} &
  \multicolumn{1}{c}{$\bar{V}_\text{LOS}$} &
  \multicolumn{1}{c}{$\sigma_{v}$} &
  \multicolumn{1}{c}{$\Sigma_0$} &
  \multicolumn{1}{c}{$r_c$} &
  \multicolumn{1}{c}{$\gamma$} &
  \multicolumn{1}{c}{$\mathcal{R}_\text{det}^2$}&
  \multicolumn{1}{c}{$R_h$} &
  \multicolumn{1}{c}{$\mathcal{M}_\text{vir}^c$}&
  \multicolumn{1}{c}{$H_Z$} &
  \multicolumn{1}{c}{$\mathcal{P}_\mathrm{relax}$}  \\

  \multicolumn{1}{c}{DCC} &
  \multicolumn{1}{c|}{AD} &
  \multicolumn{1}{c|}{JB} &
  \multicolumn{1}{c}{L} &
  \multicolumn{1}{c|}{AD} &
  \multicolumn{1}{c|}{JB} &
  \multicolumn{1}{c}{L} &
  \multicolumn{1}{c}{[km s$^{-1}$]} &
  \multicolumn{1}{c}{[km s$^{-1}$]} &
  \multicolumn{1}{c}{[gal/$h_{70}^{-2}$ Mpc$^2$]} &
  \multicolumn{1}{c}{[$h_{70}^{-1}$ Mpc]} &
  \multicolumn{1}{c}{} &
  \multicolumn{1}{c}{} &
  \multicolumn{1}{c}{[$h_{70}^{-1}$ Mpc]} &
  \multicolumn{1}{c}{[$10^{14} h_{70}^{-1}\mathcal{M}_\odot$]} &
  \multicolumn{1}{c}{[nats]}&
  \multicolumn{1}{c}{} \\
   
  \multicolumn{1}{c}{(1)} &
  \multicolumn{3}{c}{(2)} &
  \multicolumn{3}{c}{(3)} &
  \multicolumn{1}{c}{(4)} &
  \multicolumn{1}{c}{(5)} &
  \multicolumn{1}{c}{(6)} &
  \multicolumn{1}{c}{(7)} &
  \multicolumn{1}{c}{(8)} &
  \multicolumn{1}{c}{(9)} &
  \multicolumn{1}{c}{(10)} &
  \multicolumn{1}{c}{(11)} &
  \multicolumn{1}{c}{(12)} &
  \multicolumn{1}{c}{(13)} \\
\hline
001 & 1 &  1 &  0  & 0.0473 &  0.0127  &  0.1056  & 19373 &  831 &   7.12  &  2.25  &  1.42  &  0.997  & 6.48  & 40.92 & 12.98 & 0.78\\
002 & 1 &  1 &  1  & 0.0005 &  0.0010  &  0.0010  & 18319 &  697 &  17.08  &  2.83  &  3.29  &  0.995  & 4.15  & 18.40 & 13.09 & 0.78\\
003 & 0 &  0 &  0  & 0.3164 &  0.3793  &  0.1091  & 17931 &  450 &  17.57  &  1.02  &  1.31  &  0.984  & 5.71  & 10.58 & 12.17 & 0.55\\
\hline\end{tabular}
}
 \end{center}
\end{table*}

\begin{table*}
\begin{center}
\footnotesize
\textbf{Table \ref{tab:core_prop}} - \textit{continued}. 
\end{center}
\begin{center}
\resizebox{18cm}{!}{
\begin{tabular}{c ccc ccc rr rrrr crcc}
\hline
\hline
  \multicolumn{1}{c}{ID} &
  \multicolumn{3}{c}{$h_\text{test}$} &
  \multicolumn{3}{c}{$p-$values} &
  \multicolumn{1}{c}{$\bar{V}_\text{LOS}$} &
  \multicolumn{1}{c}{$\sigma_{v}$} &
  \multicolumn{1}{c}{$\Sigma_0$} &
  \multicolumn{1}{c}{$r_c$} &
  \multicolumn{1}{c}{$\gamma$} &
  \multicolumn{1}{c}{$\mathcal{R}_\text{det}^2$}&
  \multicolumn{1}{c}{$R_h$} &
  \multicolumn{1}{c}{$\mathcal{M}_\text{vir}^c$}&
  \multicolumn{1}{c}{$H_Z$} &
  \multicolumn{1}{c}{$\mathcal{P}_\mathrm{relax}$}  \\

  \multicolumn{1}{c}{DCC} &
  \multicolumn{1}{c|}{AD} &
  \multicolumn{1}{c|}{JB} &
  \multicolumn{1}{c}{L} &
  \multicolumn{1}{c|}{AD} &
  \multicolumn{1}{c|}{JB} &
  \multicolumn{1}{c}{L} &
  \multicolumn{1}{c}{[km s$^{-1}$]} &
  \multicolumn{1}{c}{[km s$^{-1}$]} &
  \multicolumn{1}{c}{[gal/$h_{70}^{-2}$ Mpc$^2$]} &
  \multicolumn{1}{c}{[$h_{70}^{-1}$ Mpc]} &
  \multicolumn{1}{c}{} &
  \multicolumn{1}{c}{} &
  \multicolumn{1}{c}{[$h_{70}^{-1}$ Mpc]} &
  \multicolumn{1}{c}{[$10^{14} h_{70}^{-1}\mathcal{M}_\odot$]} &
  \multicolumn{1}{c}{[nats]}&
  \multicolumn{1}{c}{} \\
   
  \multicolumn{1}{c}{(1)} &
  \multicolumn{3}{c}{(2)} &
  \multicolumn{3}{c}{(3)} &
  \multicolumn{1}{c}{(4)} &
  \multicolumn{1}{c}{(5)} &
  \multicolumn{1}{c}{(6)} &
  \multicolumn{1}{c}{(7)} &
  \multicolumn{1}{c}{(8)} &
  \multicolumn{1}{c}{(9)} &
  \multicolumn{1}{c}{(10)} &
  \multicolumn{1}{c}{(11)} &
  \multicolumn{1}{c}{(12)} &
  \multicolumn{1}{c}{(13)} \\
\hline
066 & 0 &  1 &  0  & 0.0852 &  0.0161  &  0.0990  & 16376 &  637 &   ---   &  ---   &  ---   &  ---    & 4.82  & 17.89 & 12.82 & 0.67\\
067 & 1 &  1 &  1  & 0.0005 &  0.0010  &  0.0010  & 14376 &  943 &  16.26  &  1.68  &  1.09  &  0.998  & 5.94  & 48.30 & 13.99 & 0.75\\
068 & 1 &  1 &  1  & 0.0005 &  0.0010  &  0.0010  & 11614 &  607 &  14.89  &  1.16  &  1.20  &  0.999  & 3.97  & 13.40 & 12.85 & 0.83\\
\hline\end{tabular}
}
\end{center}
\end{table*}

%%%%%%%%%%%%%%%%%%%%%%%%%%%%%%%%%%%%%%%%%%%%%%%%%%%%%%%%%%%%%%

\begin{table*} %flucc5
\begin{center}
\caption[Minkowski functional and shapefinders for DCC \textit{cores} (polyhedral fits).]{Minkowski functional and shapefinders for DCC \textit{cores} (polyhedral fits).}\label{tab:core_poly} 
\resizebox{13.5cm}{!}{
\begin{tabular}{crrrrrcccccr}
\hline
\hline
\multicolumn{1}{c}{ID} &
\multicolumn{1}{c}{$V$} &
\multicolumn{1}{c}{$S$} &
\multicolumn{1}{c}{$C$} &
\multicolumn{1}{c}{$\chi$} &
\multicolumn{1}{c}{$\cal{G}$}&
\multicolumn{1}{c}{$\cal{T}$}&
\multicolumn{1}{c}{$\cal{B}$}&
\multicolumn{1}{c}{$\cal{L}$}&
\multicolumn{1}{c}{$\cal{P}$}&
\multicolumn{1}{c}{$\cal{F}$}&
\multicolumn{1}{c}{$\cal{P}$/$\cal{F}$}\\

\multicolumn{1}{c}{DCC}&
\multicolumn{1}{c}{[$h_{70}^{-3}$ Mpc$^{3}$]} &
\multicolumn{1}{c}{[$h_{70}^{-2}$ Mpc$^{2}$]} &
\multicolumn{1}{c}{[$h_{70}^{-1}$ Mpc]} &
\multicolumn{1}{c}{} &
\multicolumn{1}{c}{}&
\multicolumn{1}{c}{[$h_{70}^{-1}$ Mpc]}&
\multicolumn{1}{c}{[$h_{70}^{-1}$ Mpc]}&
\multicolumn{1}{c}{[$h_{70}^{-1}$ Mpc]}&
\multicolumn{1}{c}{}&
\multicolumn{1}{c}{}&
\multicolumn{1}{c}{}\\
   
  \multicolumn{1}{c}{(1)} &
  \multicolumn{1}{c}{(2)} &
  \multicolumn{1}{c}{(3)} &
  \multicolumn{1}{c}{(4)} &
  \multicolumn{1}{c}{(5)} &
  \multicolumn{1}{c}{(6)} &
  \multicolumn{1}{c}{(7)} &
  \multicolumn{1}{c}{(8)} &
  \multicolumn{1}{c}{(9)} &
  \multicolumn{1}{c}{(10)}&
  \multicolumn{1}{c}{(11)}&
  \multicolumn{1}{c}{(12)}\\
\hline
001  &  474.5 &  567.2 &  125.0 &   1  &   0.5 &  2.50 &  4.53 &  6.63 &  0.28 &  0.18 &  1.53 \\
002  &  185.0 &  344.6 &   94.6 &  -2  &   2   &  1.61 &  2.51 &  3.63 &  0.21 &  0.18 &  1.19 \\
003  &   94.1 &  268.2 &  156.9 &  -13 &   7.5 &  1.05 &  1.46 &  1.70 &  0.16 &  0.07 &  2.18 \\
004  &   60.3 &  187.0 &  111.4 &  -3  &   2.5 &  0.96 &  1.67 &  2.53 &  0.26 &  0.20 &  1.31 \\
005  &   79.4 &  212.2 &  132.7 &   1  &   0.5 &  1.12 &  1.59 &  7.04 &  0.17 &  0.62 &  0.27 \\
\hline
\end{tabular}}
\end{center}
\end{table*}

%%%%%%%%%%%%%%%%%%%%%%%%%
\begin{table*}
\begin{center}
\footnotesize
\textbf{Table \ref{tab:core_poly}} - \textit{continued}. 
\end{center}
\begin{center}
\resizebox{13.5cm}{!}{
\begin{tabular}{crrrrrcccccr}
\hline
\hline
\multicolumn{1}{c}{ID} &
\multicolumn{1}{c}{$V$} &
\multicolumn{1}{c}{$S$} &
\multicolumn{1}{c}{$C$} &
\multicolumn{1}{c}{$\chi$} &
\multicolumn{1}{c}{$\cal{G}$}&
\multicolumn{1}{c}{$\cal{T}$}&
\multicolumn{1}{c}{$\cal{B}$}&
\multicolumn{1}{c}{$\cal{L}$}&
\multicolumn{1}{c}{$\cal{P}$}&
\multicolumn{1}{c}{$\cal{F}$}&
\multicolumn{1}{c}{$\cal{P}$/$\cal{F}$}\\

\multicolumn{1}{c}{DCC}&
\multicolumn{1}{c}{[$h_{70}^{-3}$ Mpc$^{3}$]} &
\multicolumn{1}{c}{[$h_{70}^{-2}$ Mpc$^{2}$]} &
\multicolumn{1}{c}{[$h_{70}^{-1}$ Mpc]} &
\multicolumn{1}{c}{} &
\multicolumn{1}{c}{}&
\multicolumn{1}{c}{[$h_{70}^{-1}$ Mpc]}&
\multicolumn{1}{c}{[$h_{70}^{-1}$ Mpc]}&
\multicolumn{1}{c}{[$h_{70}^{-1}$ Mpc]}&
\multicolumn{1}{c}{}&
\multicolumn{1}{c}{}&
\multicolumn{1}{c}{}\\
   
  \multicolumn{1}{c}{(1)} &
  \multicolumn{1}{c}{(2)} &
  \multicolumn{1}{c}{(3)} &
  \multicolumn{1}{c}{(4)} &
  \multicolumn{1}{c}{(5)} &
  \multicolumn{1}{c}{(6)} &
  \multicolumn{1}{c}{(7)} &
  \multicolumn{1}{c}{(8)} &
  \multicolumn{1}{c}{(9)} &
  \multicolumn{1}{c}{(10)}&
  \multicolumn{1}{c}{(11)}&
  \multicolumn{1}{c}{(12)}\\
\hline
056  &   78.7 &  244.6 &  139.0 &   2  &   0   &  0.96 &  1.76 & 11.06 &  0.29 &  0.72 &  0.40 \\
057  &   34.7 &  140.8 &  114.2 &  -2  &   2   &  0.73 &  1.23 &  3.03 &  0.24 &  0.42 &  0.59 \\
058  &  127.6 &  303.0 &   96.4 &  -6  &   4   &  1.26 &  1.53 &  3.14 &  0.09 &  0.34 &  0.28 \\
059  &   66.8 &  171.7 &   87.1 &   1  &   0.5 &  1.16 &  1.97 &  4.62 &  0.25 &  0.40 &  0.63 \\
060  &   81.3 &  162.8 &   55.5 &   1  &   0.5 &  1.49 &  2.93 &  2.94 &  0.32 &  0.00 &116.18 \\
\hline
\end{tabular}}
\end{center}
\end{table*}

%%%%%%%%%%%%%%%%%%%%%%%%%%

\begin{table*}
\begin{center}
\caption[Minkowski functional and shapefinders for DCC \textit{cores} (ellipsoidal fits).]{Minkowski functional and shapefinders for DCC \textit{cores} (ellipsoidal fits). In this fit all \textit{cores} have $\chi=2$, e.g., $\mathcal{G}=0$.}  \label{tab:core_ell} 
\resizebox{16.5cm}{!}{
\begin{tabular}{crrrrrrrcrrccc}
\hline
\hline
\multicolumn{1}{c}{ID} &
\multicolumn{1}{c}{$a$} &
\multicolumn{1}{c}{$b$} &
\multicolumn{1}{c}{$c$} &
\multicolumn{1}{c}{$\mathrm{chi}^2$} &
\multicolumn{1}{c}{$V$} &
\multicolumn{1}{c}{$S$} &
\multicolumn{1}{c}{$C$} &
\multicolumn{1}{c}{$\cal{T}$}&
\multicolumn{1}{c}{$\cal{B}$}&
\multicolumn{1}{c}{$\cal{L}$}&
\multicolumn{1}{c}{$\cal{P}$}&
\multicolumn{1}{c}{$\cal{F}$}&
\multicolumn{1}{c}{$\cal{P}$/$\cal{F}$}\\

\multicolumn{1}{c}{DCC}&
\multicolumn{1}{c}{[$h_{70}^{-1}$ Mpc]}&
\multicolumn{1}{c}{[$h_{70}^{-1}$ Mpc]}&
\multicolumn{1}{c}{[$h_{70}^{-1}$ Mpc]}&
\multicolumn{1}{c}{}&
\multicolumn{1}{c}{[$h_{70}^{-3}$ Mpc$^{3}$]} &
\multicolumn{1}{c}{[$h_{70}^{-2}$ Mpc$^{2}$]} &
\multicolumn{1}{c}{[$h_{70}^{-1}$ Mpc]} &
\multicolumn{1}{c}{[$h_{70}^{-1}$ Mpc]}&
\multicolumn{1}{c}{[$h_{70}^{-1}$ Mpc]}&
\multicolumn{1}{c}{[$h_{70}^{-1}$ Mpc]}&
\multicolumn{1}{c}{}&
\multicolumn{1}{c}{}&
\multicolumn{1}{c}{}\\
   
  \multicolumn{1}{c}{(1)} &
  \multicolumn{1}{c}{(2)} &
  \multicolumn{1}{c}{(3)} &
  \multicolumn{1}{c}{(4)} &
  \multicolumn{1}{c}{(5)} &
  \multicolumn{1}{c}{(6)} &
  \multicolumn{1}{c}{(7)} &
  \multicolumn{1}{c}{(8)} &
  \multicolumn{1}{c}{(9)} &
  \multicolumn{1}{c}{(10)}&
  \multicolumn{1}{c}{(11)}&
  \multicolumn{1}{c}{(12)}&
  \multicolumn{1}{c}{(13)}&
  \multicolumn{1}{c}{(14)}\\
\hline
001  &  5.20 &  6.37 & 12.75 &  9.4 &  1772.8 &  783.5 & 108.5 &  6.78 &  7.21 &  8.63 &  0.0306 &  0.0897 & 0.341 \\
002  &  3.67 &  5.88 & 10.09 & 11.0 &   914.3 &  517.5 &  88.6 &  5.30 &  5.83 &  7.05 &  0.0484 &  0.0941 & 0.514 \\
003  &  3.55 &  3.85 & 11.83 & 11.0 &   679.7 &  448.4 &  87.1 &  4.54 &  5.14 &  6.93 &  0.0618 &  0.1479 & 0.418 \\
004  &  2.52 &  5.95 &  8.34 & 12.2 &   525.2 &  387.8 &  76.5 &  4.06 &  5.06 &  6.09 &  0.1100 &  0.0916 & 1.200 \\
005  &  2.99 &  7.35 &  8.80 & 10.2 &   811.5 &  508.8 &  81.9 &  4.78 &  6.20 &  6.52 &  0.1295 &  0.0246 & 5.251 \\
\hline
\end{tabular}}
\end{center}
\end{table*}

%%%%%%%%%%%%%%%%%%%%%%%%%
\begin{table*}
\begin{center}
\footnotesize
\textbf{Table \ref{tab:core_ell}} - \textit{continued}. 
\end{center}
\begin{center}
\resizebox{16.5cm}{!}{
\begin{tabular}{crrrrrrrcrrccc}
\hline
\hline
\multicolumn{1}{c}{ID} &
\multicolumn{1}{c}{$a$} &
\multicolumn{1}{c}{$b$} &
\multicolumn{1}{c}{$c$} &
\multicolumn{1}{c}{$\mathrm{chi}^2$} &
\multicolumn{1}{c}{$V$} &
\multicolumn{1}{c}{$S$} &
\multicolumn{1}{c}{$C$} &
\multicolumn{1}{c}{$\cal{T}$}&
\multicolumn{1}{c}{$\cal{B}$}&
\multicolumn{1}{c}{$\cal{L}$}&
\multicolumn{1}{c}{$\cal{P}$}&
\multicolumn{1}{c}{$\cal{F}$}&
\multicolumn{1}{c}{$\cal{P}$/$\cal{F}$}\\

\multicolumn{1}{c}{DCC}&
\multicolumn{1}{c}{[$h_{70}^{-1}$ Mpc]}&
\multicolumn{1}{c}{[$h_{70}^{-1}$ Mpc]}&
\multicolumn{1}{c}{[$h_{70}^{-1}$ Mpc]}&
\multicolumn{1}{c}{}&
\multicolumn{1}{c}{[$h_{70}^{-3}$ Mpc$^{3}$]} &
\multicolumn{1}{c}{[$h_{70}^{-2}$ Mpc$^{2}$]} &
\multicolumn{1}{c}{[$h_{70}^{-1}$ Mpc]} &
\multicolumn{1}{c}{[$h_{70}^{-1}$ Mpc]}&
\multicolumn{1}{c}{[$h_{70}^{-1}$ Mpc]}&
\multicolumn{1}{c}{[$h_{70}^{-1}$ Mpc]}&
\multicolumn{1}{c}{}&
\multicolumn{1}{c}{}&
\multicolumn{1}{c}{}\\
   
  \multicolumn{1}{c}{(1)} &
  \multicolumn{1}{c}{(2)} &
  \multicolumn{1}{c}{(3)} &
  \multicolumn{1}{c}{(4)} &
  \multicolumn{1}{c}{(5)} &
  \multicolumn{1}{c}{(6)} &
  \multicolumn{1}{c}{(7)} &
  \multicolumn{1}{c}{(8)} &
  \multicolumn{1}{c}{(9)} &
  \multicolumn{1}{c}{(10)}&
  \multicolumn{1}{c}{(11)}&
  \multicolumn{1}{c}{(12)}&
  \multicolumn{1}{c}{(13)}&
  \multicolumn{1}{c}{(14)}\\
\hline
056  &  4.46 & 10.21 & 29.55 &  5.9 &  5645.3 & 2264.3 & 193.9 &  7.47 & 11.67 & 15.43 &  0.2189 &  0.1389 & 1.575 \\
057  &  2.69 &  6.61 & 12.14 &  9.0 &   907.6 &  606.8 &  96.4 &  4.48 &  6.29 &  7.67 &  0.1674 &  0.0990 & 1.690 \\
058  &  3.39 &  6.54 &  8.52 &  8.0 &   794.0 &  469.8 &  79.0 &  5.07 &  5.94 &  6.29 &  0.0791 &  0.0286 & 2.758 \\
059  &  2.06 &  5.01 &  8.04 &  8.7 &   348.3 &  308.2 &  69.3 &  3.39 &  4.44 &  5.51 &  0.1344 &  0.1079 & 1.246 \\
060  &  2.73 &  3.93 &  7.44 &  4.6 &   335.5 &  263.8 &  65.0 &  3.81 &  4.05 &  5.17 &  0.0301 &  0.1220 & 0.247 \\
\hline
\end{tabular}}
\end{center}
\end{table*}

%%%%%%%%%%%%%%%%%%%%%%%%%%%%%%%%%%%%%%%

\begin{table*}
\begin{center}
\caption[General properties of the MSCC superclusters in the sample.]{General properties of the MSCC superclusters in the sample.}
\label{tab:SCprop}
\resizebox{12cm}{!}{
\begin{tabular}{r rrr rrr}
\hline \hline
  \multicolumn{1}{c}{ID} &
  \multicolumn{1}{c}{$\mathcal{M}^\text{sc}_\text{ext}$} &
  \multicolumn{1}{c}{$V_\text{sc}$} &
  \multicolumn{1}{c}{$\rho_\text{sc}=\mathcal{M}^\text{sc}_\text{ext}/V_\text{sc}$} &
  \multicolumn{1}{c}{$\sigma_v$} &
  \multicolumn{1}{c}{$H_Z$} &
  \multicolumn{1}{c}{$\mathcal{P}_\mathrm{relax}$} \\
  
  \multicolumn{1}{c}{MSCC} &
  \multicolumn{1}{c}{[$10^{14} h_{70}^{-1} \mathcal{M}_\odot$]} &
  \multicolumn{1}{c}{[$10^3$ $h_{70}^{-3}$ Mpc$^{3}$]} &
  \multicolumn{1}{c}{[$10^{10} h_{70}^{2} \mathcal{M}_\odot$Mpc$^{-3}$]} &
  \multicolumn{1}{c}{[km s$^{-1}$]} &
  \multicolumn{1}{c}{[nats]}&
  \multicolumn{1}{c}{} \\
  
  \multicolumn{1}{c}{(1)} &
  \multicolumn{1}{c}{(2)}  &
  \multicolumn{1}{c}{(3)}  &
  \multicolumn{1}{c}{(4)}  &
  \multicolumn{1}{c}{(5)}  &
  \multicolumn{1}{c}{(6)}  &
  \multicolumn{1}{c}{(7)} \\

\hline
  1 &  143.24 &  150.4 &  9.5  &   2155 & 12.51 & 0.45 \\
 27 &   19.59 &   73.8 &  2.6  &   1159 &  9.68 & 0.53 \\
 33 &  201.42 &  268.3 &  7.5  &   1685 & 11.65 & 0.54 \\
 39 &   41.84 &   57.0 &  7.3  &    858 &  9.94 & 0.55 \\
 55 &   32.53 &   24.7 & 13.1  &   1240 & 11.45 & 0.65 \\
\hline
\end{tabular}
} 
\end{center}
\end{table*}

%%%%%%%%%%%%%%%%%%%%%%%%%%%%%%%%%%%%%%%%%%%%%%%%%%%
\section*{Acknowledgments}
The authors are grateful
for the discussions and suggestions of Dr. Varun Sahni and Dr. Satadru
Bag that helped improve the volume estimates of structures through polyhedral
surface fitting.

This research was supported by CONAHCyT through
a PhD grant and Universidad de Guanajuato (DAIP) CIIC-0162/22 and CIIC-0088/24 projects grants. H. A. benefited from grant CIIC 211/2024 of Universidad de Guanajuato.

%%%%%%%%%%%%%%%%%%%%%%%%%%%%%%%%%%%%%%%%%%%%%%%%%%

\section*{Data Availability}

Data resulting from the present work are available
in the manuscript, both printed and digitally, and codes can be requested to the corresponding author.



%\bibliographystyle{mnras}


%%%%%%%%%%%%%%%%%%%%%%%%%%%%%%%%%%%%%%%%%%%%%%%%%%

%%%%%%%%%%%%%%%%% APPENDICES %%%%%%%%%%%%%%%%%%%%%

%%%%%%%%%%%%%%%%%%%%%%%%%%%%%%%%%%%%%%%%%%%%%%%%%%

% Don't change these lines
\bsp	% typesetting comment
\label{lastpage}
\end{document}